\documentclass{uri}
\usepackage{amsmath}
  \usepackage{paralist}
  \usepackage{graphics} %% add this and next lines if pictures should be in esp format
  \usepackage{epsfig} %For pictures: screened artwork should be set up with an 85 or 100 line screen
\usepackage{graphicx}  \usepackage{epstopdf}%This is to transfer .eps figure to .pdf figure; please compile your paper using PDFLeTex or PDFTeXify.
\usepackage{tikz}
\usetikzlibrary{arrows.meta} % for control over arrows
\usetikzlibrary{math} % for arithmetic computations
\usetikzlibrary{decorations.pathreplacing,decorations.pathmorphing} % for interface diagram

 \usepackage[colorlinks=true]{hyperref}
 \usepackage{eso-pic}
\usepackage[percent]{overpic}
\usepackage{subcaption}
\usepackage{transparent}
\usepackage{color}
\definecolor{redcolor}{rgb}{1.0,0,0}
\definecolor{bluecolor}{rgb}{0,0,1.0}
\definecolor{browncolor}{rgb}{.6,.2,.2}

\hypersetup{urlcolor=blue, citecolor=red}

\def\currentvolume{X}
 \def\currentissue{X}
  \def\currentyear{200X}
   \def\currentmonth{XX}
    \def\ppages{X--XX}

\newtheorem{example}{Example}

\theoremstyle{definition}

\newcommand {\uu}  { {\bf u} }

\newcommand {\xx}  { {\bf x} }

\newcommand {\yy}  { {\bf y} }

\newcommand {\FF}  { {\bf F} }

\newcommand{\R}{\mathbb{R}}

\newcommand {\qq}  { {\bf q} }
\newcommand {\pp}  { {\bf p} }

\newcommand {\bv}  { {\bf v} }
\newcommand {\vv}  { {\bf v} }
\newcommand {\ww}  { {\bf w} }
\newcommand {\ff}  { {\bf f} }
\newcommand {\gb}  { {\bf g} }
\newcommand {\av}  { {\bf a} }

\newcommand {\cc}  { {\bf c} }
\newcommand {\dd}  { {\bf d} }

\newcommand {\zero}  { {\bf 0} }
\newcommand {\blambda}  { {\boldsymbol \lambda} }

\def\nml{\mathsf{\scriptscriptstyle N}}
\def\tng{\mathsf{\scriptscriptstyle T}}

\def\Jac{J_C}
\def\vc{\bar{\vv}}
\def\BB{B}
\def\TT{T}
\def\nl{ {\boldsymbol\eta }}
\def\nli{ \tilde{\boldsymbol\eta }}
\def\ContactDom{\mathcal{C}} % Contact domain (points on the boundary)
\DeclareMathOperator*{\argmax}{argmax}

\def\ext{\mathrm{ext}}

\def\c{\mathrm{c}}
\def\f{\mathrm{f}}

\newcommand{\Description}[2][]{}

\definecolor{turquoise}{cmyk}{0.65,0,0.1,0.1}
\definecolor{purple}{rgb}{0.65,0,0.65}
\definecolor{dark_green}{rgb}{0, 0.5, 0}
\definecolor{orange}{rgb}{0.8, 0.6, 0.2}
\definecolor{red}{rgb}{0.8, 0.2, 0.2}
\definecolor{brown}{rgb}{0.5, 0.16, 0.16}
\newcommand{\uri}[1]{{\color{black}#1}}
\newcommand{\urio}[1]{{\color{black}#1}}
\newcommand{\rev}[1]{{\color{black}#1}}
\newcommand{\egorrev}[1]{{\color{black}#1}}
\newcommand{\pai}[1]{{\color{black}#1}}
\title[SIMULATING DEFORMABLE OBJECTS FOR ANIMATION] %Use the shortened version of the full title
      {Simulating deformable objects for computer  animation:  a numerical  perspective}

\author[Uri M. Ascher,  Egor Larionov, Seung Heon Sheen, and Dinesh K. Pai]{}


 \email{ascher@cs.ubc.ca}
 \email{egor@cs.ubc.ca}
 \email{heonsheen@gmail.com}
 \email{pai@cs.ubc.ca}

\thanks{The first and last authors are supported by NSERC Discovery grants 84306 and RGPIN/2017-04604 respectively. Pai's research was also supported by a Canada Research Chair and an NSERC Idea-to-Innovation grant co-sponsored by Vital Mechanics.}

\begin{document}
\maketitle

% Enter the first author's name and address:
\centerline{\scshape Uri M. Ascher$^*$ and Egor Larionov and Seung Heon Sheen and Dinesh K. Pai}
\medskip
{\footnotesize
% please put the address of the first author
 \centerline{Dept. Computer Science}
   \centerline{University of British Columbia}
   \centerline{Vancouver, BC, V6T 1Z4, Canada}
} % Do not forget to end the {\footnotesize by the sign }

%\medskip

%\centerline{\scshape Egor Larionov and Dinesh K. Pai}
%last-name3}
%\medskip
%{\footnotesize
 % please put the address of the second  and third author
% \centerline{ First line of the address of the second author}
 %  \centerline{Other lines}
 %  \centerline{Springfield, MO 65810, USA}
%}

\bigskip

% The name of the associate editor will be entered by an editorial staff
% "Communicated by the associate editor name" is not needed for special issue.
 %\centerline{(Communicated by the associate editor name)}

%The abstract of your paper
\begin{abstract}
We examine a variety of numerical methods that arise when considering dynamical systems in the context of
physics-based simulations of deformable objects.
Such problems  arise in various applications, including animation, robotics, control and fabrication.
The goals and merits of suitable numerical algorithms \pai{for these applications} are different from  those of typical numerical analysis research in dynamical systems.
\pai{Here the mathematical model is not fixed {\em a priori} but must be adjusted as necessary to capture the desired behaviour, with an emphasis %is 
on effectively producing lively animations of objects with complex geometries.
Results are often judged by how realistic they appear to observers (by the ``eye-norm'') as well as by the efficacy of the numerical procedures employed.}
And yet, we show that with an adjusted view numerical analysis and applied mathematics can contribute  significantly
to the development of appropriate methods and their analysis in a variety of areas including finite element methods, stiff and highly oscillatory ODEs,
model reduction, and constrained optimization.
\end{abstract}

%%%%%%%%%%%%%%%%%%%%%%%%%%%%%%%%%%%%%%%%%%%%%%%%%%%%%%%%%%%

\section{Introduction}
\label{sec:introduction}

\begin{figure}[htp]
\begin{center}
  % Requires \usepackage{graphicx}
  % replace aims_logo.eps by your figure file name
  \includegraphics[width=2.04in]{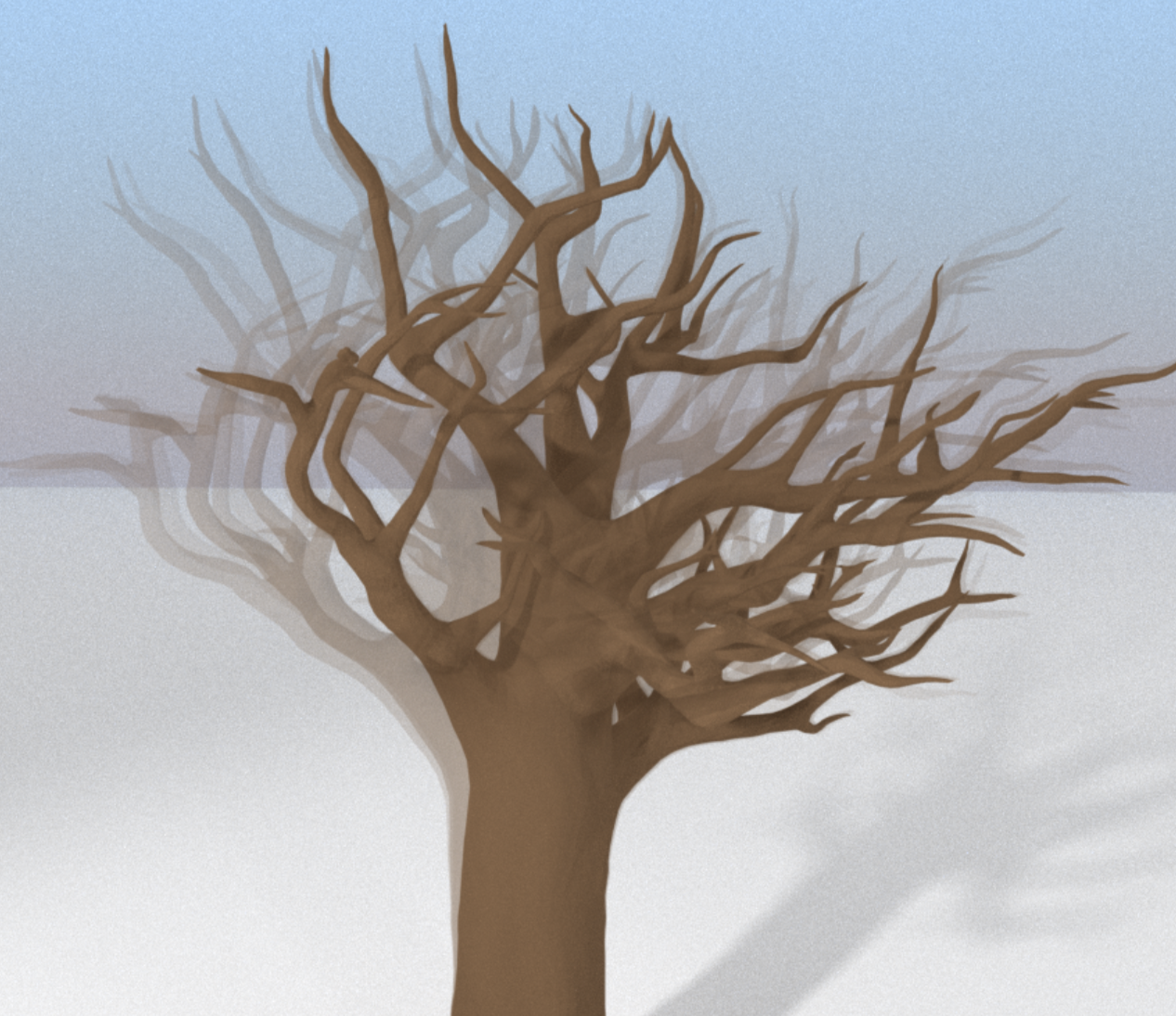}
    \includegraphics[width=1.756in]{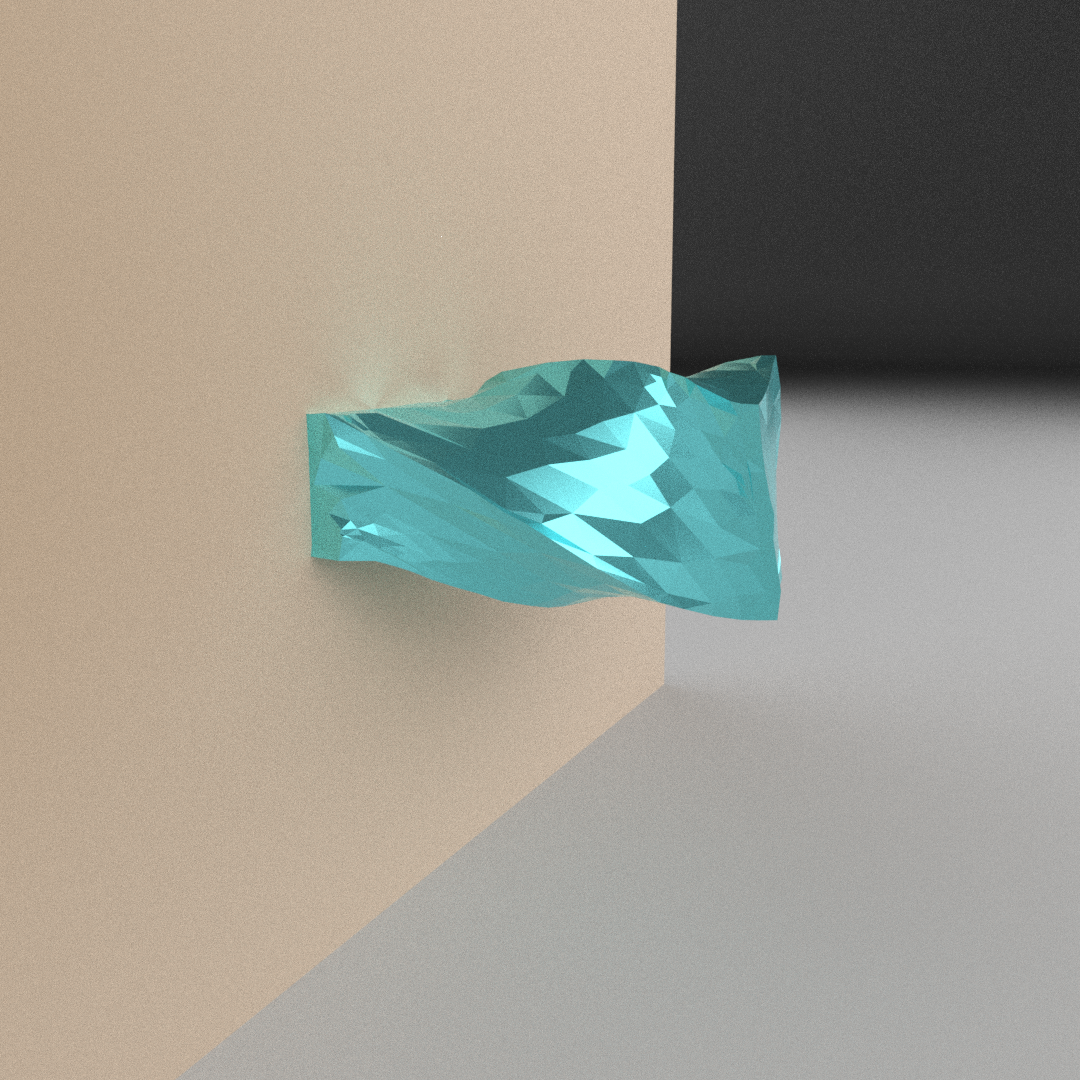}\\
  \caption{Deformable articulated objects: a swaying tree and a constrained jelly brick; cf. \cite{Chen:siere:2020}}\label{fig:objects}
  \end{center}
\end{figure}

Physics-based simulations of deformable objects are ubiquitous in computer graphics today.
They arise in various applications including 
animation, robotics, control and fabrication.
Whereas in typical numerical analysis research of dynamical systems the mathematical model is given
and the task is to construct and prove the properties of computational methods that efficiently and reliably 
provide highly accurate solutions in often idealized circumstances, here the emphasis is typically on effectively producing lively-looking
animations of objects with rather complex geometries (see Figure~\ref{fig:objects}). The quality of the results is judged by the ``eye-norm'' as well as
the efficacy of the numerical procedures employed.
Numerical analysts and applied mathematicians thus can contribute  significantly while adjusting to such a different environment,
as the usual methods of scientific computing remain important and relevant and yet are not always ideal for a given task.

\begin{example}
To demonstrate the issues involved, suppose that we have discretized in space the equations of motion for one of the objects
in Figures~\ref{fig:objects} and~\ref{fig:FEMmesh} using a tetrahedral finite element mesh. Assuming only linear elastic forces $\ff$, the equations of motion in time  $t$ read
\begin{eqnarray}
 M \ddot \qq = \ff (\qq ) = -K\qq.
 \label{1ex}
 \end{eqnarray}
 Here $\qq (t)$ are the node coordinates of the finite elements, the corresponding accelerations are denoted by $\ddot \qq$, $K$ is the stiffness matrix,
  and $M$ is the mass matrix.
 Assume that both of these matrices are constant and symmetric positive definite (SPD).
 Note that $\qq$ is not a function of the mesh: rather, it consists of the mesh nodes; see Figure~\ref{fig:FEMmesh}.
 \begin{figure}[htp]
\begin{center}
  % Requires \usepackage{graphicx}
  % replace aims_logo.eps by your figure file name
  \includegraphics[width=2.0in]{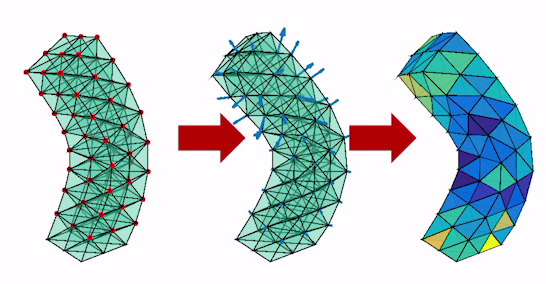}\\
     \caption{Moving tetrahedral FEM mesh for position coordinates $\qq(t)$.}\label{fig:FEMmesh}
  \end{center}
\end{figure}
 
 The matrix $K$ may well have large eigenvalues (\pai{which usually correspond to high frequency oscillations}), both because upon using a fine spatial discretization higher and higher modes are captured
 and because the object's material itself may be stiff (expressed by a large  Young's modulus). 
 
 %The system \eqref{1ex} is 
 In~\eqref{1ex} we have a simple Hamiltonian system, and thus typical suitable time discretizations that may come to mind
 are conservative: they would be symmetric and non-dissipative
 (such as the trapezoidal rule), 
 if not downright symplectic~\cite{hlw}.
 
 However, there is always some damping in the motion of such objects. Moreover, symplectic and symmetric time discretization methods cannot be L-stable~\cite{hw}, and
 much care is therefore required near boundary constraints such as contact~\cite{chen2017}.
 \urio{In applied mathematics literature} the damping force has traditionally been handled by adding to $\ff$ a Rayleigh damping force term 
 \begin{eqnarray}
 \ff_{\rm dmp} (\qq,\dot \qq) = (\alpha K + \beta M)\dot \qq ,
 \label{2ex}
 \end{eqnarray}
%  to $\ff$,
 where $\alpha > 0$ and $\beta \geq 0$ are suitable constants. However, the Rayleigh damping, although theoretically rather helpful, does not
 cover all manners of motion behaviour  observed in practice.
 \pai{Even though there has been work in measuring material properties of real objects \cite{Pai+01} and even humans \cite{Pai2018human}}, the selection of  suitable %\pai{friction} and 
 damping parameters
 (as is the choice of parameters in \eqref{2ex})
 in different \urio{practical} circumstances remains a challenging problem that is often solved by trial and error (e.g., \cite{xuba}).
 
 In the computer graphics literature, in contrast to the above,  the semi-implicit backward-Euler (SI) method of Baraff and Witkin~\cite{baraff1998} 
has been widely employed. 
 This method allows for stable simulations even when large time steps are used for efficiency reasons,
and it is very stable when incorporating contacts and collisions due to its heavy damping
and error localization properties.
Moreover, numerical damping is in agreement with the observation that our visual system does not
detect high frequency vibrations, even when objects with large Young's modulus are simulated.

And yet, recent years have also seen growing concerns that heavy numerical damping    
may be unsuitable for many purposes, because it can only be controlled via the time step size, so it
does not distinguish phenomena related to material heterogeneity and more complex damping forces~\cite{Chung1993,chen2017exponential}.
Animations produced using such methods often do not appear to be sufficiently lively.
Use of the SI method in applications such as control and fabrication, \urio{where more faithfulness to physics is desired}, has also been considered debatable~\cite{chen2017}.

A further concern that arises when the force $\ff$ is nonlinear is that SI has long been known to occasionally diverge wildly 
where the fully implicit backward Euler (BE)
still yields acceptable results.
This ushers %in 
another issue seldom considered seriously in the geometric numerical integration community, namely, finding solution methods
for the nonlinear systems of algebraic equations that typically arise in the present context. \fbox
 
\label{examp1}
\end{example}

In this paper we consider several issues that involve the transportation of methods and ideas from the numerical community 
to the present computer animation context. Following a short discussion in Section~\ref{sec:model} 
of the equations of motion that arise from an FEM semi-discretization, 
we present and analyze in Section~\ref{sec:siere} several suitable methods for the stiff dynamical systems that arise.
These include difference methods  and  exponential ones, as well as additive methods that involve both through a model reduction algorithm.
Numerical experiments with soft objects are reported in Section~\ref{sec:soft}, and methods for handling contact and friction
constraints are considered in Section \ref{sec:contact}. These involve several interesting modelling and optimization issues.
Conclusions are offered in Section~\ref{sec:conclusion}.

\rev{Virtually all research papers on the present topic in the computer graphics and robotics literature are supplemented
by video clips that demonstrate object {\em motion} obtained by the proposed numerical methods.
Such material can usually be accessed through authors' project pages, home pages, or even YouTube on the web.
Our own relevant videos for \cite{chen2017exponential,Chen:siere:2020,Chen:coarsen:2019} and \cite{larionov21}, 
respectively, can be found at \\ 
https://www.youtube.com/watch?v=oRGuC9GMm8w \\
https://www.cs.ubc.ca/$\sim$ascher/papers/csap.m4v \\
https://www.youtube.com/watch?v=aIAKBsT96to\&feature=youtu.be \\
https://elrnv.com/vid/fcses\_tog2021.webm  }

%%%%%%%%%%%%%%%%%%%%%%%%%%%%%%%%%%%%

%%%%%%%%%%%%%%%%%%%%%%%%%%%%%%%%%%%%%%%%%%%%%%%%%%%%%%%%%%%

\section{Motion of deformable objects}
\label{sec:model}

A common approach for discretizing an elastodynamic system is to first apply a finite element method (FEM) 
in space, {\em already at the variational level}.
%see Figure~\ref{fig_mesh}. 
%%\uri{[This stolen figure will have to be replaced!]}
%\begin{figure}
%  \centering   	
%%  \begin{overpic}[width=\columnwidth]{swinging_arma_cost}
% %    \put(15,30){\includegraphics[width=0.5\columnwidth]{swinging_arma0}}  
%%  \end{overpic}
%\includegraphics[width=0.45\columnwidth]{figures/eiff_simple_mesh.pdf}
%\caption{A spatial tetrahydral mesh.}
%\label{fig_mesh}
%\end{figure}
%%
Usually in physics-based animations linear element basis functions are used on tetrahedra (or less frequently, hexahedra),
although there are works that employ higher degree elements \cite{Longva2020}.
The short course by Sifakis and Barbic \cite{sifakis12} is a good introduction for this material and much more.
See \cite{ciarlet1988three} for an exhaustive mathematical treatment of elasticity.
 
This spatial discretization results in a large ordinary differential equation (ODE) system in time $t$, {written in standard notation as}
\begin{subequations}
\begin{eqnarray}
M\ddot{\qq} = {\ff}_{\rm tot}(\qq,\bv) , \label{1a}
\end{eqnarray}
where the unknowns $\qq = \qq(t)$ are nodal displacements in the FEM mesh with  corresponding {nodal} velocities $\bv (t) = \dot \qq (t)$.
The mass matrix $M$ is SPD {and sparse}. The total force is further written as
\begin{eqnarray}
\ff_{\rm tot} = {\ff}_{\rm els}(\qq) + {\ff}_{\rm dmp} (\qq,\bv) + {\ff}_{\rm con}(\qq,\vv) + {\ff}_{\rm ext}(\qq) , \label{1b}
\end{eqnarray}
where ${\ff}_{\rm els}(\qq) $ are elastic forces, ${\ff}_{\rm dmp} (\qq,\bv)$ are damping forces (e.g., as given in \eqref{2ex}),
${\ff}_{\rm con}$ are forces due to contact and friction constraints, and $ {\ff}_{\rm ext}$ are external forces such as gravity.
\label{1}
\end{subequations}

\rev{The choice of spatial FEM discretization generally affects the resulting animations obtained by any subsequent
numerical treatment of the semi-discretization~\eqref{1}, see for instance~\cite{chen2017,Chen:coarsen:2019}.
Nonetheless, unlike in some shock wave calculations, there is enough separation between space and time here to allow
a separate treatment of~\eqref{1}, and we proceed to do this following almost all relevant literature.} 

The elastic and damping forces are written as
\[{\ff}_{\rm els}(\qq (t)) = -\frac{\partial }{\partial \qq} W(\qq(t)),\quad {\ff}_{\rm dmp}(\qq (t),\bv (t)) = -D\bv(t) ,\]
where \(W(\qq(t))\) is the elastic potential of the corresponding hyperelasticity model. 
%For example, \(W(q(t))\) is quadratic in the linear elasticity model, 
%and it is quartic in the linear FEM model with StVK material. 
%In other models such as co-rotated FEM with linear material, neo-Hookean material, 
%or artificially designed material~\cite{xu15}, nonlinear potentials are required to describe the desired physical behavior.
We further define the tangent {\em stiffness matrix} \(K= -\frac{\partial}{\partial \qq} \ff_{\rm els}(\qq (t)) .\)
This matrix is constant and SPD if $\ff_{\rm els}$ is linear, but for nonlinear elastic forces
it depends on the unknown $\qq(t)$ and may occasionally become indefinite.
Note that  $K$ is possibly very large in size: fortunately, it is also rather sparse.

Typical nonlinear elastic forces include neo-Hookean, StVK and Mooney-Rivlin~\cite{ciarlet1988three}.
Nonlinear force effects in computer graphics include in addition as-rigid-as-possible (ARAP) forces \cite{sorkine:2007}, artist-generated forces,
and co-rotated FEM, which is an FEM specialty not typically known to general FEM experts! (See, e.g.,   \cite{wwyalh}.)
 
The damping matrix $D = D(\qq (t))$ is symmetric nonnegative definite at all times, as in \eqref{2ex};
see, for instance, \cite{barbic05,ciarlet1988three,sifakis12,chen2017,chen2017exponential}.

A similar-looking problem arises upon using a mass-spring system~\cite{baraff1998,boxerman2004decomposing}, 
though the spectrum of the corresponding stiffness matrix
is less challenging to split in that case.

The ODE system in~\eqref{1} can be written as a first-order system
\begin{equation}\label{eq:FirstOrder}
\dot{\uu}(t) \equiv
\begin{bmatrix}
\dot{\qq}\\  
\dot{\bv} 
\end{bmatrix}  = 
\begin{bmatrix}
{0} & I\\
-M^{-1}K & -M^{-1}D
\end{bmatrix}
\begin{bmatrix}
\qq\\
\bv
\end{bmatrix}
+
\begin{bmatrix}
\bf{0}\\
{\cc}\end{bmatrix} \equiv {\bf F} (\uu)  .
\end{equation}%
The split form of writing these equations expresses an expectation that the remainder term $\cc =  M^{-1}( \ff_{\rm tot} + K\qq  + D\vv )$
is somehow dominated by the first term.
Let us denote the (often large) length of $\uu$ by $n$. In the tree example of Figure~\ref{fig:objects}, $n \approx 180,000$.
In industrial applications $n$ can get to millions.

\rev{Note that we could have avoided $M^{-1}$ in \eqref{eq:FirstOrder} by considering momenta $\pp = M\bv$.
However, in the application domain considered here the mass matrix is often diagonal and simple, and the velocities
are of direct interest.}

%%%%%%%%%%%%%%%%%%%%%%%%%%%%%%%%%%%%%%%%%%%%%%%%%%%%%%%%%%%

\section{Efficient simulation methods for stiff objects}
\label{sec:siere}

In this section we mention and discuss several favourite \rev{time} discretization methods that have seen use in practice.
See also the recent~\cite{Loschner2020}.

\subsection{Finite difference methods}
\label{sec:findif}

Our prototype ODE problem~\eqref{eq:FirstOrder}  must be discretized before simulation.
Let us concentrate on one time interval, %where we 
stepping from time $t=t_0$ where we know an approximate solution $\uu_0 \approx \uu (t_0)$ 
to the next time level $t_1 = t_0 + h$ where the unknown approximation $\uu_1 \approx \uu(t_1)$ is sought.

\rev{The time step size $h$ is typically large, and it is kept constant as the simulation proceeds.
This is in contrast to canned initial value ODE software, where usually the step size is adapted at each step, chosen so as to satisfy
an equidistribution of a local error estimate~\cite{apbook,hw}. But here, the time step often has a physical meaning and is tied
to a constant rate of operation of some physical apparatus, especially in robotics. Moreover, the meaning of ``accuracy'' is more
relaxed in the present context compared to what numerical analysts are used to.}  

For instance, the backward Euler (BE) method is given by
\begin{eqnarray}
\uu_1 = \uu_0 + h  {\bf F} (\uu_1). \label{3.1.1}
\end{eqnarray}
A large nonlinear system of algebraic equations must be solved for $\uu_1$ when employing an implicit method such as~\eqref{3.1.1}.
For this purpose Newton's method for finding a root of
\[ \gb (\uu_1) = \uu_{1} - h{\bf F} (\uu_1) - \uu_0 = \zero\]
decrees iteration to convergence of
\[ \uu_{1}^{(s+1)} = \uu_{1}^{(s)} - [I - h J(\uu_{1}^{(s)}) ]^{-1}  (\gb (\uu_1^\rev{{(s)}})) , \quad s = 0, 1, \ldots \]
where the {\em Jacobian} matrix $J = \frac{\partial \FF}{\partial \uu}$ is large and sparse.
This iteration may require careful control and possible modification in practice.

The SI method is \uri{derived by} %obtained upon 
applying to BE just one Newton iteration starting at $\uu_{1}^{(0)} = \uu_0$, obtaining
 \begin{eqnarray}
 \uu_{1} = \uu_0 + h [I  - h J(\uu_0) ]^{-1}  \FF (\uu_0) .  \label{3.1.2}
 \end{eqnarray}
For a linear ODE, BE and SI are the same; but for nonlinear forces and large step sizes $h$ they can differ significantly.

In the sequel we retain the notation SI for the semi-implicit version of BE.
But of course a semi-implicit version can be obtained in the same fashion for any implicit difference method.
We will distinguish such methods by adding the letter 'S' in front of the method's name.
For instance, the BDF2 method is a two-step method that requires also the past approximation $\uu_{-1}$ at $t_0-h$.
This method reads
 \begin{eqnarray}
\uu_1 = \uu_0 + \frac 13 \left( \uu_0 -  \uu_{-1} + 2h  {\bf F} (\uu_1) \right), \label{3.1.3}
\end{eqnarray}
and its semi-implicit version is denoted by SBDF2.

As stated above, 
the popular SI method and even implicit integrators such as BE and 
higher order backward differentiation formulae (BDF) introduce significant, step-size dependent, artificial damping.
{All BDF methods collocate the ODE system only at the unknown time level}. %~\cite{ascher08}. 
This typically yields 
a potentially heavy damping of high frequency modes that
has often been observed and used to advantage in practice~\cite{chen2017exponential,Chen:siere:2020}.

Let us next \pai{discuss} two diagonally implicit Runge-Kutta (DIRK) methods~\uri{\cite{alexander77,hw,leveque2}}.
The TR-BDF2 method is motivated by wanting to use BDF2 over just one time step rather than two, making it a one-step method.
But for this, a value for the solution at $t_0+h/2$ is required. This approximation is obtained using the trapezoidal rule
(which by itself introduces no artificial damping), giving the method

\begin{subequations}
\begin{eqnarray}
\uu_{1/2} &=& \uu_0 + \frac h4 \left( \FF (\uu_{1/2}) +  \FF (\uu_0 ) \right) \label{10a} \\
\uu_{1} &=& \uu_0 + \frac 43 ( \uu_{1/2} - \uu_0 )  + \frac h3 {\FF}(\uu_{1}) . \label{10b}
\end{eqnarray}
\label{10}
\end{subequations}

This method performed well in comparative experiments reported in~\cite{Loschner2020}. It
has two implicit stages, each requiring the solution of a nonlinear system of size $n$.
S-versions for these are derived directly as before. In fully implicit $r$-stage RK methods a nonlinear system of size $rn$ must be solved,
and this is avoided here by the DIRK method.
The matrices to be inverted (or rather, decomposed) here, similarly to \eqref{3.1.2},  are $I -  \frac h4 J$ and $I -  \frac h3 J$.
A nearby singly DIRK (SDIRK, where 'S' here stands for 'singly', not 'semi') method uses the trapezoidal rule to advance to $t_0+\gamma h$ first, 
where $\gamma = 2 - \sqrt{2}$.
This yields the method
\begin{subequations}
\begin{eqnarray}
\uu_{\gamma} &=& \uu_0 + \frac {\gamma h}2 \left( \FF (\uu_{\gamma}) +  \FF (\uu _0) \right) \label{11a} \\
\uu_{1} &=& \uu_0 + \frac {2\beta}\gamma ( \uu_{\gamma} - \uu_0 )   + \frac{\gamma h}2 {\FF}(\uu_{1}) , \label{11b}
\end{eqnarray}
where $\beta = \frac{\sqrt{2}}4$.
\label{11}
\end{subequations}

\uri{The corresponding Butcher Tableau representation for the methods~\eqref{10} and \eqref{11}  is
\begin{eqnarray*}
\begin{array}{r|rrr}
0 &  &  &  \\
1/2 & 1/4 & 1/4 & \\
1 & 1/3 & 1/3  & 1/3 \\
\hline
 & 1/3 & 1/3 &  1/3
\end{array} \quad {\rm ~~and~~} \quad 
\begin{array}{r|rrr}
0 &  &  &  \\
\gamma & \frac{\gamma}2 &  \frac{\gamma}2 & \\
1 & \beta & \beta  &  \frac{\gamma}2   \\
\hline
 & \beta & \beta&   \frac{\gamma}2
\end{array}
\end{eqnarray*}
respectively. In the sequel we will refer for simplicity  to~\eqref{11} as {\em the} SDIRK method:  there is no other SDIRK method in this paper
to get confused by.}

The advantage of the SDIRK method \eqref{11}, favoured in the initial value ODE literature~\cite{alexander77,hw,butcher2000}, 
%over \eqref{10} 
is that %it is SDIRK, not just DIRK, and thus 
it can be implemented somewhat more efficiently, because
upon freezing the Jacobian matrix \uri{$J$}  over the step's  interval, we get to invert the same matrix $I - \frac{\gamma h}2 J$
en route to solving the two nonlinear problems in the stages of \eqref{11}. The same LU decomposition may then be used for both stages.
This advantage fades away if an iterative solution method must be employed for this large problem.
The order of accuracy in both stages of these methods equals $2$, and they can both be shown to be L-stable~\cite{hw,butcher2000,leveque2,apbook}.
For \eqref{10}, in particular, upon considering the test equation $\dot u = \lambda u$ for a complex scalar $\lambda$ and writing the resulting step
as $u_1 = R(z ) u_0$ with $z = \lambda h$, it is not difficult to show that  
\[ R(z) \approx 5/z \rightarrow 0 \quad {\rm as~} |z| \rightarrow \infty .\]
L-stability follows.

On the other hand, for very stiff problems both of these DIRK variants share a disadvantage, as the following example shows.
\begin{example}
Consider the test equation
\[ \dot u = -1000 u ,\quad  u(0) = 1, \]
and use a step of size $h=.1$, say, to advance the first step (i.e., $t_0 = 0$). 
Applying the BDF2 method to this yields $u_{1} >  0$ that is 
close to $0$ but still nonnegative. However,
the trapezoidal method in both \eqref{10a} and \eqref{11a}
would yield a negative $u_{\gamma } \approx -1$.

Now consider the ODE system for $\uu (t) = (u,w)^T$, given by
\[ \dot u = -1000 u, \;\; \dot w = \log (u) . \]
The BDF2 method will complete the step successfully, whereas both
featured DIRK methods will get stuck, being unable to evaluate $\log (u_{\gamma})$ for calculating $w_{1}$. 

For mildly stiff problems, typically bad effects such as instability accumulate over {\em several} time steps. 
However, in very stiff problems like %such as 
the one under consideration here, %however,
a method such as trapezoidal could already cause irreparable inaccuracy over just half a %one 
step, thus not allowing the BDF2 that follows a 
chance to fix things up. \fbox

\label{examp2}
\end{example}

\subsection{Damping curves}
\label{sec:damp}

The test equation used to draw stability regions and define concepts such as A-stability and L-stability \cite{hw,apbook}
is less natural when it comes to analyzing the second-order problem in Example~\ref{examp1} or more generally the system \eqref{1}.
Instead we consider the scalar ODE
\begin{eqnarray}
\ddot q + \omega^2q = 0, \label{testn}
\end{eqnarray}
where $\omega$ is a potentially large frequency. See~\cite{chen2017exponential,Chen:siere:2020,chhu}. %\cite{ashu18}

Rewriting \eqref{testn} as a first order system and applying a difference discretization such as
any of the methods described above, we can write the operation over one time step as
\[ \uu_{1} = T\uu_0 , \]
where $T$ is a  constant transformation matrix (of size $2 \times 2$ for a one-step method).
The spectral radius of $T$, $\rho (T)$, then corresponds to $|R(z)|$ above, as it gives the contraction factor in $\| \uu \|$.\footnote{
\urio{Throughout this paper we use the notation $\| \cdot \|$ for the $\ell_2$-norm.}}
Of course, in \eqref{testn} the oscillations are undamped, so we expect damping numerical methods such as all of the above
to approximate more closely the solution of the modified ODE
\begin{eqnarray}
\ddot q + d^{\rm method} \dot q + \omega^2q = 0, \label{testm}
\end{eqnarray}
where the damping coefficent $d^{\rm method} \geq 0$ depends on $\omega$ and $h$ in such a way that $d^{\rm method}/\omega$
depends only on the product $\omega h$.
The (simple) procedure for determining $d^{\rm method}$ is described in Section 3.1 of \cite{chen2017exponential}.
%and Section 2.1 of \cite{ashu18}.

Of course, for conservative discretizations such as trapezoidal, implicit midpoint, collocation at Gaussian points, and Kane's method~\cite{kmow} 
there is no artificial damping and $d^{\rm method} = 0$. But for damping methods we can obtain telling tales in this fashion.
Figure~18 in \cite{Chen:siere:2020} shows that the BE damping curve $d^{\rm BE}$ has a similar shape as $d^{\rm BDF2}$,
 but the damping is roughly twice as large in magnitude.
An interesting comparison of the DIRK methods and BDF2 (all 2nd order methods) is presented in Figure~\ref{fig:damp1} below. 
\begin{figure}[htp]
\begin{center}
  \includegraphics[width=5.0in]{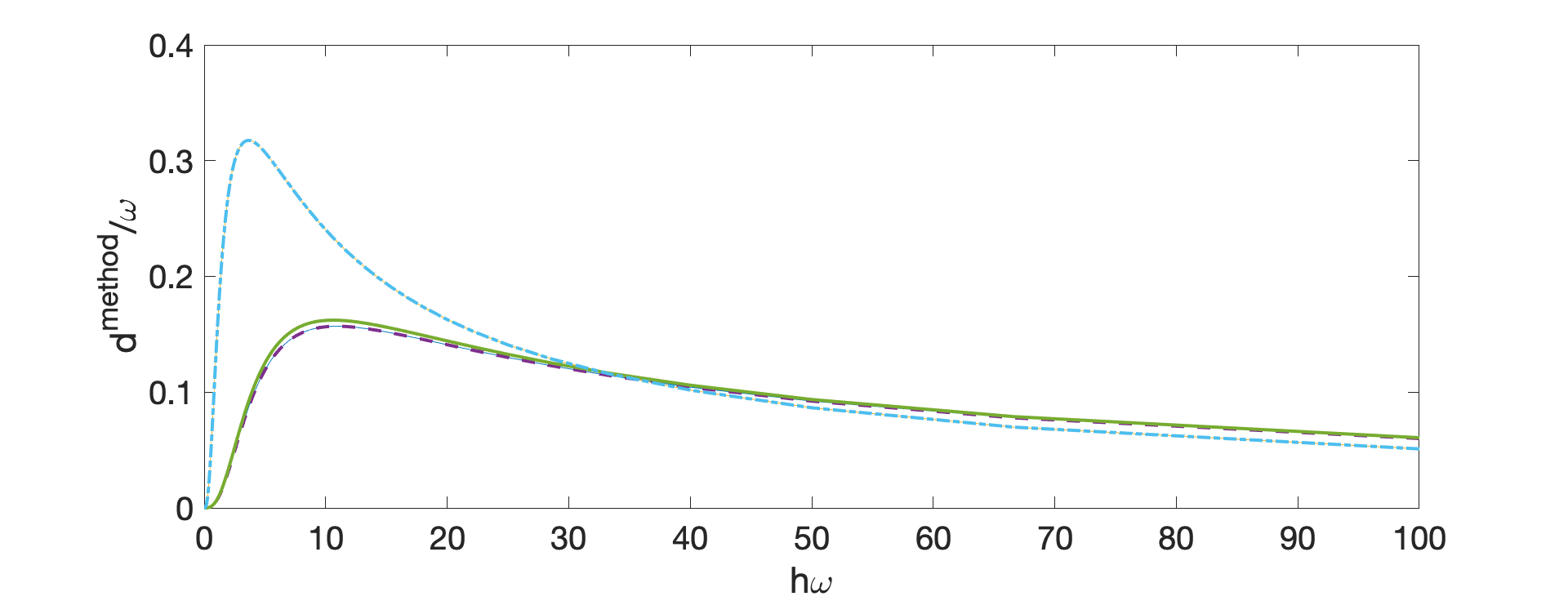}
  \caption{Damping curves for the SDIRK method (solid line), TR-BDF2 (dashed) and BDF2 (dash-dot).
  The two DIRK methods behave similarly, and they differ significantly from BDF2.}\label{fig:damp1}
  \end{center}
\end{figure}
Clearly, the two DIRK methods are close to each other in this respect as well, with TR-BDF2~\eqref{10} damping slightly less.
The BDF2 curve, on the other hand, suggests significantly more damping around $h \omega \approx 1$ and less when $h\omega \gg 1$.
The DIRK curves are more appealing, in fact, implying less dependence on the frequency as well as less damping for low frequencies. 

\subsection{Solving the nonlinear algebraic equations}
\label{sec:nonlinear}

Curiously, in situations where there is a real need to solve the nonlinear equations arising from implicit discretizations
of the problem \eqref{1}, or \eqref{eq:FirstOrder}, this may become a challenge.
Among the reasons are the sheer size of the problem, the fact that large steps are considered so $\uu_0$ may not be a close approximation
to $\uu_{1}$, and the observation that we are considering situations where the semi-implicit version (i.e., one Newton iteration) is not sufficient.

A general approach for solving systems of nonlinear equations is to cast this as an optimization problem; \rev{cf.~\cite{Gast:2015}}.
We may always consider minimizing a scaled $\ell_2$-norm of the residual.
For instance, applied to BE \eqref{3.1.1} this reads
\begin{eqnarray}
\min_{\uu_{1}}  \phi_{LSBE}  = \frac 12 \| B  (\uu_{1}  -   \uu_0 - h{\bf F} (\uu_1 )) \|  , \label{opt2}
\end{eqnarray}
where the scaling matrix $B$ multiplying the residuals in this expression may depend on the time step.
It can be the identity, or a diagonal matrix that scales the rows,
or even the inverse of a Jacobian evaluated once at the beginning of the time step. 

The advantages of such formulations are that they are general in terms of the forces regardless of integrability,
and that constraints may be added directly to form a constrained optimization problem.
A disadvantage, however, is that the gradient expression involves $J^TJ$, and this reduces the sparsity of the Jacobian
and squares its condition number. 

An alternative option is to employ a damped Newton strategy, using~\eqref{opt2} (and its likes for other implicit discretizations)
only as a merit function. %; see~\cite{amr}.
But let us stay with the ``lazy option'' of just employing  an optimization software package for our present problem.

To simplify matters, let us consider just elastic forces in \eqref{1b}. 
This force is integrable, and we have $\ff_{\rm tot} = -\frac {\partial W (\qq)}{\partial \qq}$ with $W$ the potential energy.
Thus we can write the nonlinear equations $\uu_{1} =   \uu_0 + h{\bf F}(\uu_{1})$ for BE
as the necessary condition for a minimum of
\begin{eqnarray}
\min_{\vv_{1}}  \phi_{BE}  &\equiv& \frac 12 \|\vv_1 - \vv_0 \|_M^2  + W (\qq_{1})  , \label{optbe} \\
{\rm where}~~\qq_{1} &=& \qq_0 + h \vv_{1} . \nonumber
\end{eqnarray}
See, e.g.,~\cite{li20}.
Here we have used the ``energy norm'' $\| \vv \|_M = \sqrt{\vv^TM\vv}$ for the kinetic energy term.

A similar, if slightly more complex, expression can be written for BDF2:
The nonlinear equations $\uu_{1} =   \uu_0 + \frac 13 \left( \uu_0 - \uu_{-1} + 2h{\bf F}(\uu_{1}) \right)$ are written as the necessary condition
for a minimum of
\begin{eqnarray}
\min_{\vv_{1}}  \phi_{BDF2}  &\equiv& \frac 12 \|\vv_{1} - \tilde \vv\|_ M^2    +  W (\qq_{1})  , \label{optbdf2} \\
{\rm where}~~ \qq_{1} &=& \qq_0 + \frac 13 \left ( \qq_0 - \qq_{-1} + 2h \vv_{1} \right), \nonumber \\
{\rm with}~~ \tilde \vv &=& \vv_0 + \frac 13 (\vv_0 - \vv_{-1}) . \nonumber
\end{eqnarray}

Method-dependent optimization problems can be similarly constructed also for other discretization methods,
including trapezoidal and the DIRK methods.
An advantage in these objective functions is that no matrix sparsity is sacrificed. 
Other integrable forces such as gravity and %certain contact and 
\uri{frictionless contact} forces can be handled in a similar way: see Section~\ref{sec:contact}.
An explicitly stated damping force can be a bit more tricky and require some approximation such as freezing $D$.

\subsection{Newmark methods}
\label{sec:Newmark} 

Newmark methods are very popular in structural mechanics and related fields, less so in the numerical ODE community.
{\em Generalized $\alpha$} is one such family of methods that has been used in computer graphics applications.
It has a knob (parameter) to control the amount
of artificial diffusion per frequency~\cite{Chung1993}.
These methods are one-step and $2$nd order accurate.

Let us rewrite the ODE \eqref{1}  as a simple semi-explicit differential-algebraic equation (DAE):
\begin{eqnarray}
\dot \qq = \vv, \quad \dot \vv = \av, \quad \zero = M\av - \ff_{\rm tot} (\qq,\vv) .
\label{21}
\end{eqnarray}
%The method has coefficients to play with: $\alpha_m \neq 1, \alpha_f, \beta$, and $\gamma$.
%Let $\alpha = \alpha_m - \alpha_f$. 
The time step unknowns from $t_0$ to $t_{1}$
are $\qq_{1}, \vv_{1}$ and $\av_{\nu}$, i.e., it is a staggered time stepping
for the acceleration, where $\nu$ is a parameter.
%See Section~2.1 of~\cite{ashu18} as well as
See~\cite{chen2017exponential} for further details, including damping curves for the generalized $\alpha$ method.
\uri{These curves are comparable to those of} a simple mixture of the trapezoidal and BDF2 methods.
Our experiments,
%there and below, 
as well as those of others, have not left us with the feeling \pai{that} there is much additional uplift to
look for here, although these methods are solid performers in general.

\subsection{Exponential methods}
\label{sec:exponential}

Instead of finite difference discretizations as described above, one can resort to exponential methods.
Such methods are more accurate when they work, and they approximate the entire modal spectrum
with little artificial damping (indeed, $d^{\rm method} \equiv 0$ in the modified test equation~\eqref{testm}). 
Moreover, these methods are semi-implicit rather than fully implicit, and require no solutions of nonlinear algebraic equations!
  
Several exponential integration schemes have been proposed in the computer graphics 
literature; see~\cite{michels016,michels2017,chen2017exponential} and references therein.
In principle they all involve calculations of the action of the matrix exponential of $M^{-1}K$ on various vectors.
%If the tangent stiffness matrix is known to be always positive definite, especially in case it is constant,
%then methods using its square root matrix may be devised.
%However, for nonlinear forces that appear in practice there is the need to exponentiate at every time step, since {$K$
%depends on $\qq(t)$},
%and there is no such guarantee of positive definiteness.

\subsubsection{ERE}

The exponential Rosenbrock Euler (ERE) method of~\cite{chen2017exponential} is particularly suitable for nonlinear forces in 
\uri{some} %such 
challenging situations. Its nominal order
is commensurate with that of the {BE} 
%backward Euler (BE) 
method, although it is generally much more accurate than BE.

%This method is defined as follows. 
With the Jacobian $J$ defined as before we first write~\eqref{eq:FirstOrder} as
\begin{eqnarray}
 \dot \uu = \FF (\uu ) =  J\uu+ {\bf c}(\uu),\; \quad {\rm where} \;\; J = \frac{\partial \FF}{\partial \uu}, \label{ere1}
  \end{eqnarray}%
so ${\bf c}(\uu)$ is nonlinear but hopefully can be treated as an inhomogeneity. 
Next, write the step from $t_0$ to $t_{1}$ as
approximating the integral form of \eqref{ere1} over one step
\begin{equation*}
\uu (t_1)=\text{exp}(h J)\uu_0 + \int_{t_0}^{t_{1}}\text{exp}((t_{1}-s)J){\bf c}(\uu(s))\text{d}s .
%\uu(t)=\text{exp}(t J)\uu(0)+\int_{0}^{t}\text{exp}((t-s)J)\vec{c}(\uu(s))\text{d}s
\end{equation*}

This yields the ERE step
\begin{subequations}
\begin{eqnarray}
\uu_{1}  &=& \uu_0 + h \phi_{1}(h J)\FF(\uu_0),  \quad {\rm with} \label{eq:ERE1} \\
\phi_{1}(Z) &=& Z^{-1}(\exp(Z) - I) . \label{phi_1}
\end{eqnarray}
\label{eqERE}
\end{subequations}

The remaining key issue is the evaluation of the product of $\phi_1 (hJ)$ times a vector.
Of special concern is the exponential matrix $\exp (hJ)$, which is very large and unfortunately no longer sparse.
This is discussed in detail in~\cite{chen2017exponential}.
A Krylov-space method is employed there for the implementation used  for approximating the product of $\exp (hJ)$ with 
vectors ${\bf F}$~\cite{moler2003survey,niesen2012,al11}.

\subsubsection{SIERE}
\label{sec:SIERE}

The method used in ~\cite{chen2017exponential} for calculating $\exp (hJ)\FF$ is efficient for objects made of material that is not very stiff. 
%The explicit formation of the exponential matrix appearing in $\phi_1$ is out of the question for large matrices $J$ such as we have in the ERE method
%for Eq.~\eqref{eq:FirstOrder}, as this matrix become full (dense) even though $K$, $M$ and $D$ are all rather sparse. 
%Researchers resort instead to Krylov {subspace} methods. 
However, {as shown in Figure~\ref{ere_cost}, % \uri{Figure Y1}
% and Table~\ref{tab:time},
these methods become expensive for stiff problems, \uri{because the required number of Krylov vectors can become very large.
Thus, %limiting 
the utility of exponential integrators in the context of stiff deformable object simulations is limited.} 
%The root of this problem 
%comes from 
%{is in} the fact that the required number of Krylov vectors  
%. {needed} in order to fully resolve the error for a stiff system  as in Eq.~\eqref{eq:FirstOrder} with a wide spectrum matrix and large \({\bf F}(\uu)\) 
%{can become very large}. One might consider choosing a smaller time step to assist the matrix function evaluation for stiff objects; however, in 
%{physics-based simulations to stay competitive we expect to be able to keep employing}
%computer graphics we generally like to  keep 
%large time steps independently of the material parameters. 
The method proposed in~\cite{Chen:siere:2020} %described below
%in Section~\ref{sec:method} 
alleviates this difficulty by applying \uri{ERE}
%the exponential integrator 
only in a suitable subspace,
and the exponential matrix {evaluation} at each time step is drastically simplified due to matrix diagonalization.
\begin{figure}
\centering
\begin{overpic}[width=0.5\textwidth, trim=0 0 15 0, clip]{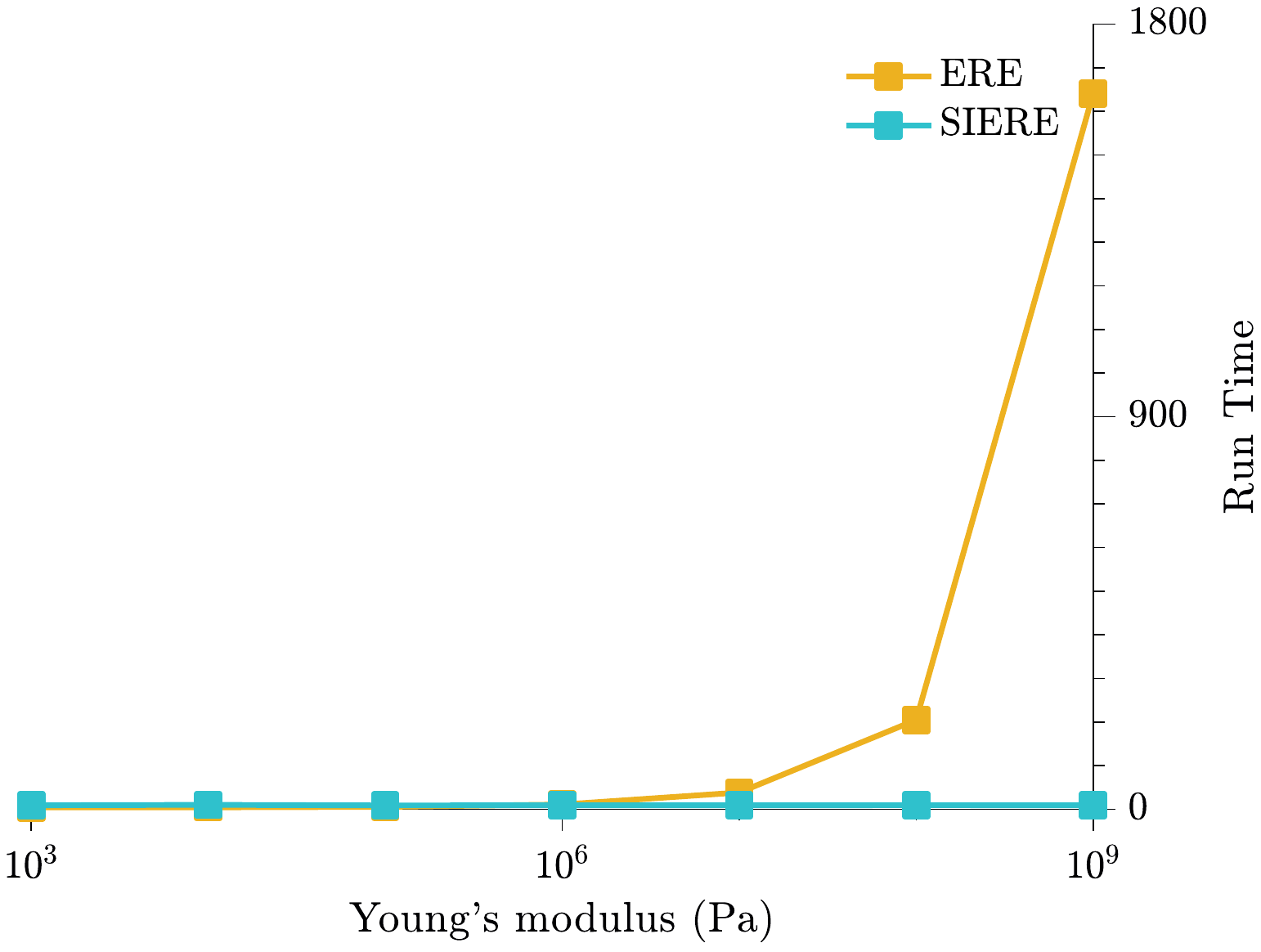}
     \put(10,30){\includegraphics[width=0.15\textwidth]{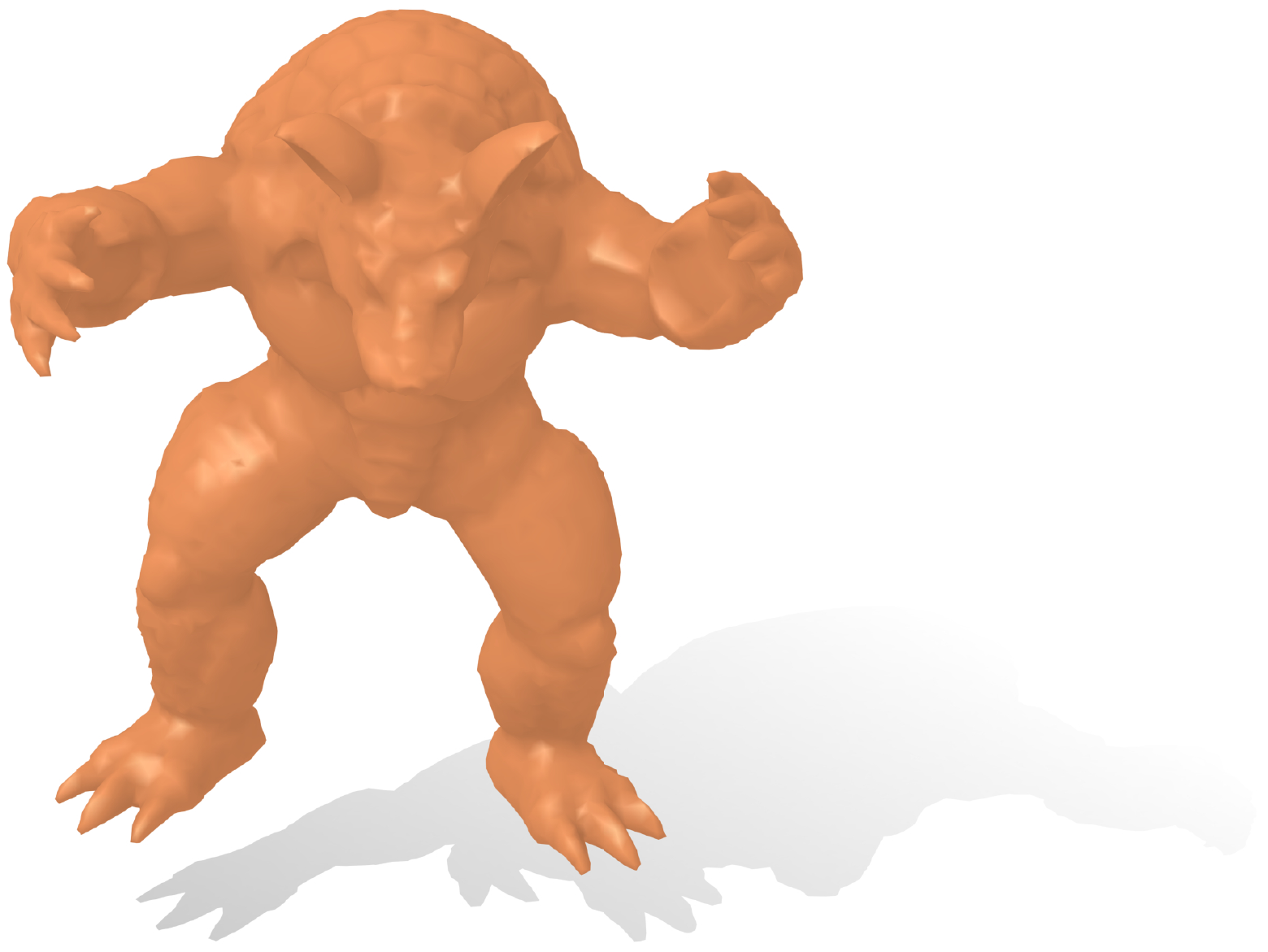}}
     \put(100,42){\rotatebox[origin=c]{90}{\footnotesize Run Time (seconds)}}
  \end{overpic}
  \caption{Computational costs for a swinging armadillo simulation~\cite{Chen:siere:2020}.
  The cost of exponential integrators including ERE %can  better preserve the oscillations, but at a prohibitive cost 
  becomes prohibitive as the stiffness parameter %and/or system size 
  increases.  By contrast, the cost of SIERE does not grow significantly with stiffness.
  \label{ere_cost}}
%  \end{center}
\end{figure}

Observe at first that the advantages and disadvantages of ERE and SI are largely complementary.
%On one hand, 
SI is cheap and stable, %with respect to collisions and boundaries, 
and its efficiency does not deteriorate for highly stiff problems;
 yet it is lethargic and loses energy rapidly, and it may exhibit divergence for large steps when BE is not approximated well.
 On the other hand,
 ERE approximates all modes reasonably well, and it is also semi-implicit;
 but \uri{its performance deteriorates %becomes very expensive 
 for very stiff problems.} % and may experience instabilities in high modes then.
 So, SIERE combines these two methods %(as the name implies) 
 such that ERE works on the \pai{$s$ lowest} modes 
 while SI is applied to dampen the higher modes. Typically, $1 \leq s \leq 20$, with $s = 5$ being already suitable for many practical applications.
 
We employ the {\em additive method} framework:
If ${\bf F}(\uu)$ can be written for each time instance $t$ as a sum of two parts
\begin{eqnarray}
{\bf F}(\uu (t))  = {\bf H}(\uu (t))+{\bf G}(\uu (t)) , \label{eq:splitting-form}
\end{eqnarray}
\uri{then we can apply different integration schemes for ${\bf H}$ and for ${\bf G}$}.
%\uri{then it is possible to 
%write $\uu(t)=\xx(t)+\yy(t)$, where $\xx$ and $\yy$ satisfy
%\begin{subequations}
%\begin{gather}
%\dot{\xx}={\bf H}(\uu)={\bf H}(\xx+\yy),\label{eq:ode-for-ere}\\
%\dot{\yy}={\bf G}(\uu)={\bf G}(\xx+\yy).\label{eq:ode-for-be}
%\end{gather}
%We can therefore  apply one integration scheme to Eq.~\eqref{eq:ode-for-ere} and another one to Eq.~\eqref{eq:ode-for-be},
%add them up and 
%obtain a combined integration scheme for our ODE system~\eqref{eq:FirstOrder}.}
%~\cite{Luan2017,ascher08}.
%\end{subequations}
%In the present context, 
Combining BE and ERE in this way gives the method BEERE, defined by
\begin{eqnarray}
\uu_{1} = \uu_0 + h {\bf H} (\uu_{1}) + h \phi_1 (h J_G){\bf G} (\uu_0) ,  \label{beere}
\end{eqnarray}
while SIERE is defined upon approximating  BE by its semi-implicit iteration
\begin{equation}
    \uu_{1} = \uu_0 + (I-hJ_{H})^{-1}(h {\bf H}(\uu_0) + h \phi_1 (hJ_{G}) {\bf G}(\uu_0)) . \label{SIERE}
\end{equation} 

We can of course combine other methods with ERE this way. For instance, the BDF2ERE method is defined by
\begin{subequations}
\begin{eqnarray}
\hat \uu &=& \frac 13 \left( 4\uu_0 - \uu_{-1} +  2h\phi_{1}(hJ_{G}){\bf G}(\uu_0) \right) , \label{bdf2erea} \\
\uu_{1} &=&  \hat \uu + \frac {2h}3 {\bf H}(\uu_{1}),  \label{bdf2ereb}
\end{eqnarray}
\label{BDF2ERE}
\end{subequations}
and its semi-implicit version is
\begin{eqnarray}
\uu_{1} &=&  \hat \uu + \frac {2h}3 (  {\bf H}(\uu_0) + J_H(\uu_1 - \uu_0))   \nonumber \\
&=& \uu_0 +  \frac 13 (I -   \frac {2h}3 J_H)^{-1} \left( \uu_0 - \uu_{-1} + 2h{\bf H}(\uu_0)  + 2h \phi_{1}(hJ_{G}){\bf G}(\uu_0)  \right)  .     \label{sbdf2ere}
\end{eqnarray}
The %obtained  
SBDF2ERE method given in \eqref{sbdf2ere} is only slightly more expensive than SIERE.
% yet it has the potential of being more expressive and fail less often than SIERE. 

\rev{The above methods from \eqref{beere} to \eqref{sbdf2ere} can be naturally viewed as {\em splitting methods} as well; see, e.g., \cite{ascherbook}.}

%\uri{[Explain next how the decomposition is done, how low rank corrections are handled, and what are the pitfalls.]} $~$
%We now describe how 
The splitting of $\FF(\uu)$ as in~\eqref{eq:splitting-form} is achieved in~\cite{Chen:siere:2020}
using a model reduction technique. Assuming 
that the dominant part of the force $\ff_{\rm tot}$ is in $-M^{-1}K \qq$, 
the Fourier ansatz $\qq (t) = \xx \exp (\imath \sqrt {\lambda}t )$ at each step start $t=t_0$ gives the
generalized eigenvalue problem
\[    K (\qq (t_0)) \xx = \lambda M \xx  .\]%
%\item
With $X_s$ the (long and skinny) matrix of first $s$ eigenmodes and {$D_s = {\rm diag}~ [\lambda_1, \ldots , \lambda_s]$}
%(in {\sc Matlab} use {\tt eigs}) 
we obtain
\[ KX_s = MX_sD_s  . \]}%

Next, define
\begin{eqnarray*}
& & {\bf G} (\uu)  =  \begin{bmatrix} \vv_{G} \\ M^{-1}\ff_{G} \end{bmatrix} \quad 
{\bf H} (\uu) = \begin{bmatrix} \vv_H \\ M^{-1}\ff_{H} \end{bmatrix}, 
%\quad {\rm where}   \cr
%& &  {\vv_{G} = X_{s}X_{s}^{T}M\vv} \quad
%\vv_{H}  =  \vv-\vv_{G} \nonumber \\ 
%& & {\ff_{G}  =  MX_{s}X_{s}^{T}\ff}  \quad
%\ff_{H}  =  \ff-\ff_{G} .
\end{eqnarray*}
\noindent where ${\vv_{G} = X_{s}X_{s}^{T}M\vv}, \ 
\vv_{H}  =  \vv-\vv_{G}, \  
{\ff_{G}  =  MX_{s}X_{s}^{T}\ff}, \
\ff_{H}  =  \ff-\ff_{G}$.
{It follows that
\begin{eqnarray}
 J_{G} = 
\begin{bmatrix}0 & X_{s}X_{s}^{T}M\\
-X_{s}X_{s}^{T}KX_{s}X_{s}^{T}M & 0 \end{bmatrix}  \quad 
J_{H} =
%\left[\begin{array}{cc} 
\begin{bmatrix}
0 & I\\
-M^{-1}K & 0
\end{bmatrix} - J_{G}. \label{24} 
\end{eqnarray}%
For the ERE update in the subspace we write %for SIERE
\begin{subequations}
\begin{eqnarray}
    \uu_{1} = \uu_0 + h(I- h J_{H})^{-1}\left({\bf H} (\uu_0)
    +  \begin{bmatrix}
X_s & 0\\
0 & X_s
\end{bmatrix}
    \phi_1 (h J^r_{G}) {\bf G}^r(\uu_0)\right) , \label{25a}
\end{eqnarray}
%with the \(2s \times 2s\) matrix
where
\begin{eqnarray}
J^r_{G} = 
    \begin{bmatrix}
    0 & I\\
    -X_s^TKX_s & 0
    \end{bmatrix} \quad
%    \end{eqnarray}
 %   and the  \(2s\times2n\) matrix
 %   \begin{eqnarray}
{\bf G}^r(\uu) = 
    \begin{bmatrix}
    X_s^T M \vv\\
    X_s^T \ff
    \end{bmatrix} . \label{25b}
\end{eqnarray}
Note that the ERE expression {$h \phi_{1}(h J_{G}){\bf G}(\uu_0)$} can be
trivially evaluated in the subspace first and then projected back to the original
space.
\label{25}
\end{subequations}%
}%

Sifting through the suddenly detailed expressions above, we note the crucial point that in \eqref{25a} the very large
linear algebraic system that must be solved involves the matrix $I - h J_{H}$ which is no longer sparse.
Fortunately, this matrix is an $s$-rank correction of the sparse matrix $I - hJ$
\begin{eqnarray}
I- h J_{H} = (I - h J)  
+h \left(Y_1 Z_1^T + Y_2 Z_2^T\right), \label{26}
\end{eqnarray}%
where {$J$} is a typically sparse FEM matrix  and
$Y_i$ and $Z_i$ are skinny $n \times s$ matrices. 
This saves the day, in principle: 
\begin{itemize}
\item
For iterative methods (e.g., conjugate gradient) evaluating\\ 
$J_{H} \ww = J\ww + Y_1 \left( Z_1^T\ww) + Y_2 (Z_2^T\ww\right) $ 
for any given vector $\ww$ is efficient.
\item
For direct methods (i.e., variants of Gaussian elimination) we employ the celebrated {Sherman-Morrison-Woodbury (SMW)} formula
{\begin{eqnarray*}
\left(A + YZ^T \right)^{-1} = A^{-1} - A^{-1} Y \left(I + Z^TA^{-1}Y\right)^{-1}Z^TA^{-1} \label{smw}
\end{eqnarray*}}%
for $A = I - h J$.
\end{itemize}
See~\cite{Chen:siere:2020} for more details and extensive demonstration of the efficacy of SIERE. 

Other hybrid methods such as BEERE and BDF2RE require the solution of nonlinear algebraic equations at each time step.
Note that \eqref{bdf2ereb} looks just like BE, and such is the case also for the nonlinear part of \eqref{beere}.
Hence the optimization methods decribed in Section~\ref{sec:findif}, in particular \eqref{optbe}, can be adapted to include BEERE
and BDF2ERE. There is a catch, though, in that the procedures for evaluating $(I - J_H)\ww$ and $(I-J_H)^{-1}\ww$ for
any given vector $\ww$ are specialized, so care must be exercised when using canned optimization software.
\rev{We feel that the exciting development here is rather in the semi-implicit variants SIERE and SBDF2ERE.
For instance, with an appropriate choice of $s$, the semi-implicit expression for $\hat \uu$ in~\eqref{bdf2erea} often gives
a sufficiently good approximation for $\uu_1$ so that one subsequent Newton iteration for \eqref{bdf2ereb}, i.e., SBDF2ERE, suffices!}

\section{Efficient simulation methods for soft objects}
\label{sec:soft}

%\uri{[Heon examples?]}

%\input{softbody_revise.tex}
Soft objects,  {such as the characters that appear in animation movies aimed at young audiences, commonly  arise in computer graphics applications.} 
Here, the DIRK methods introduced in Section~\ref{sec:findif}, namely
TR-BDF2 and SDIRK, %appear to 
have a good record~\cite{xuba,Loschner2020}, \rev{even though occasionally an additional stabilization or damping may be required.} 

For soft objects with nonlinear elastic forces, there may be large deformations during a single time step that change the eigenvalues {and modes described above}
significantly, enough to cause the eigenmodes to cross~\cite{Chen:coarsen:2019}. 
If contemplating use of additive methods such as SIERE, then for stiff objects it is sufficient to perform the spectral decomposition only once, at the beginning.
% when the problem is stiff, since there are much smaller deformations with much less changes to the eigenvalue over the timeframe.
However, for soft objects, one \rev{may} need to perform the spectral decomposition
%compute the eigenvalues much 
more frequently, possibly \rev{even} at every frame (i.e., time step interval). 
%Computing the eigenvalues more frequently 
Doing this does produce more \rev{stable}  %lively 
results {in our experiments}, 
but at the cost of increased runtime, since computing the first $s$ eigenpairs is significantly more expensive than the other parts of the integrator within one time frame.
\rev{Such additional cost is still small compared to just assembling the forces and the stiffness matrix at each time step, though.}
This raises the question of how frequently one should compute this decomposition without sacrificing too much compute time while
%the eigenvalues so that the eigenmodes are 
still remaining faithful to the simulated deformation. 
%without sacrificing much compute time.

\begin{figure}[htp]
	\begin{center}
		\includegraphics[width=4in]{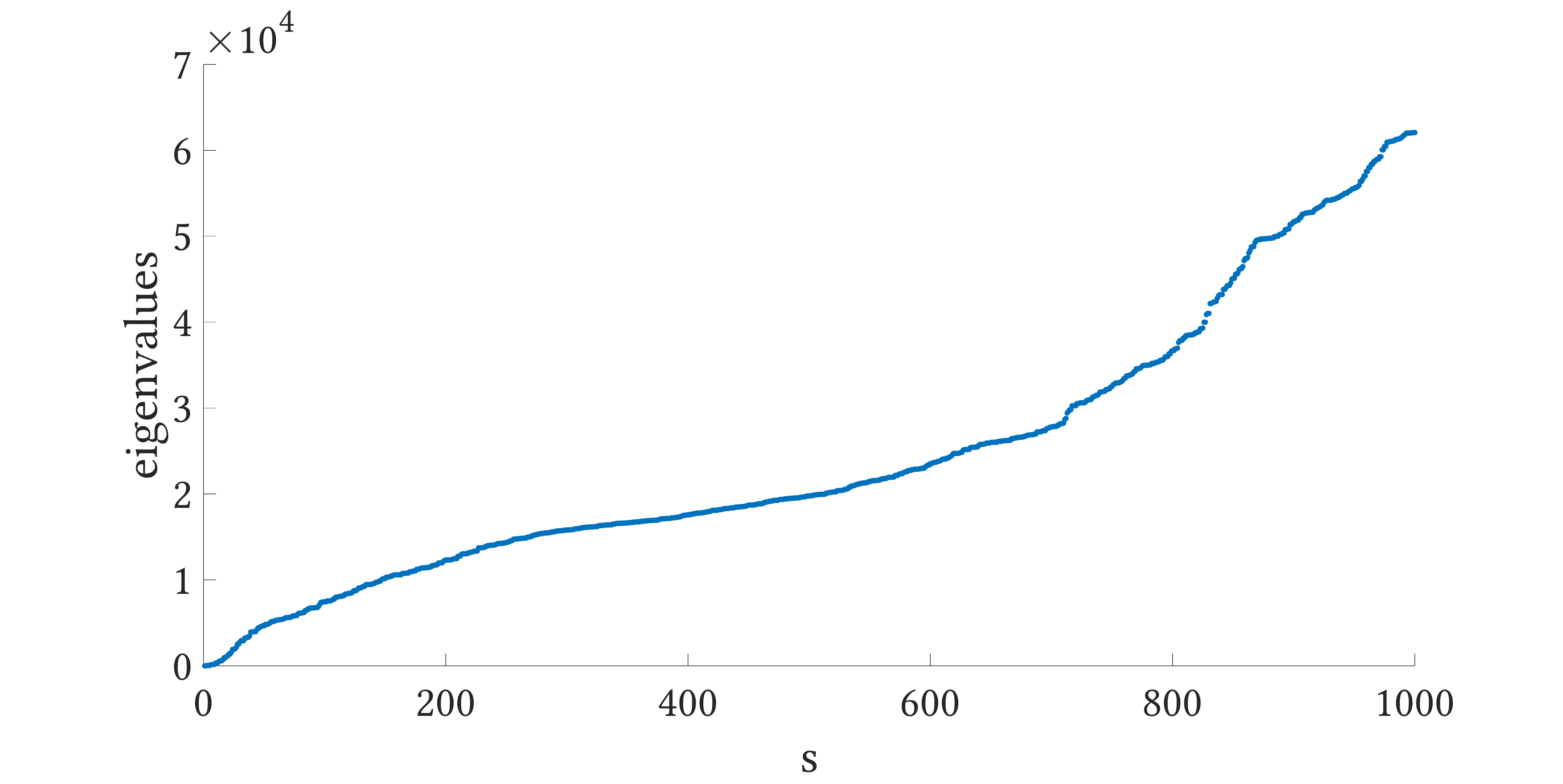}
		\caption{Plot of of the first 1000 eigenvalues of a soft body problem. 
			\label{fig:soft_eigs}}
	\end{center}
\end{figure}

Another issue is that for soft problems it can be hard to clearly distinguish the first $s$ low frequency modes from higher ones for a small $s$.
%For example, we plot in 
Consider Figure~\ref{fig:soft_eigs}, where we plot the first $1000$ eigenvalues for a soft body problem. % is plotted, where 
Here, there is no clear separation in the eigenvalues, and it is hard to clearly distinguish a small %few set 
number of lowest eigenmodes that dominate the dynamics.
In other words, more eigenmodes play a significant role in the visible dynamics. 
% which implies that the artificial damping for integrating $\HH(\uu)$ is also crucial.

\begin{figure}[htp]
	\centering
	\begin{overpic}[width=5.0in]{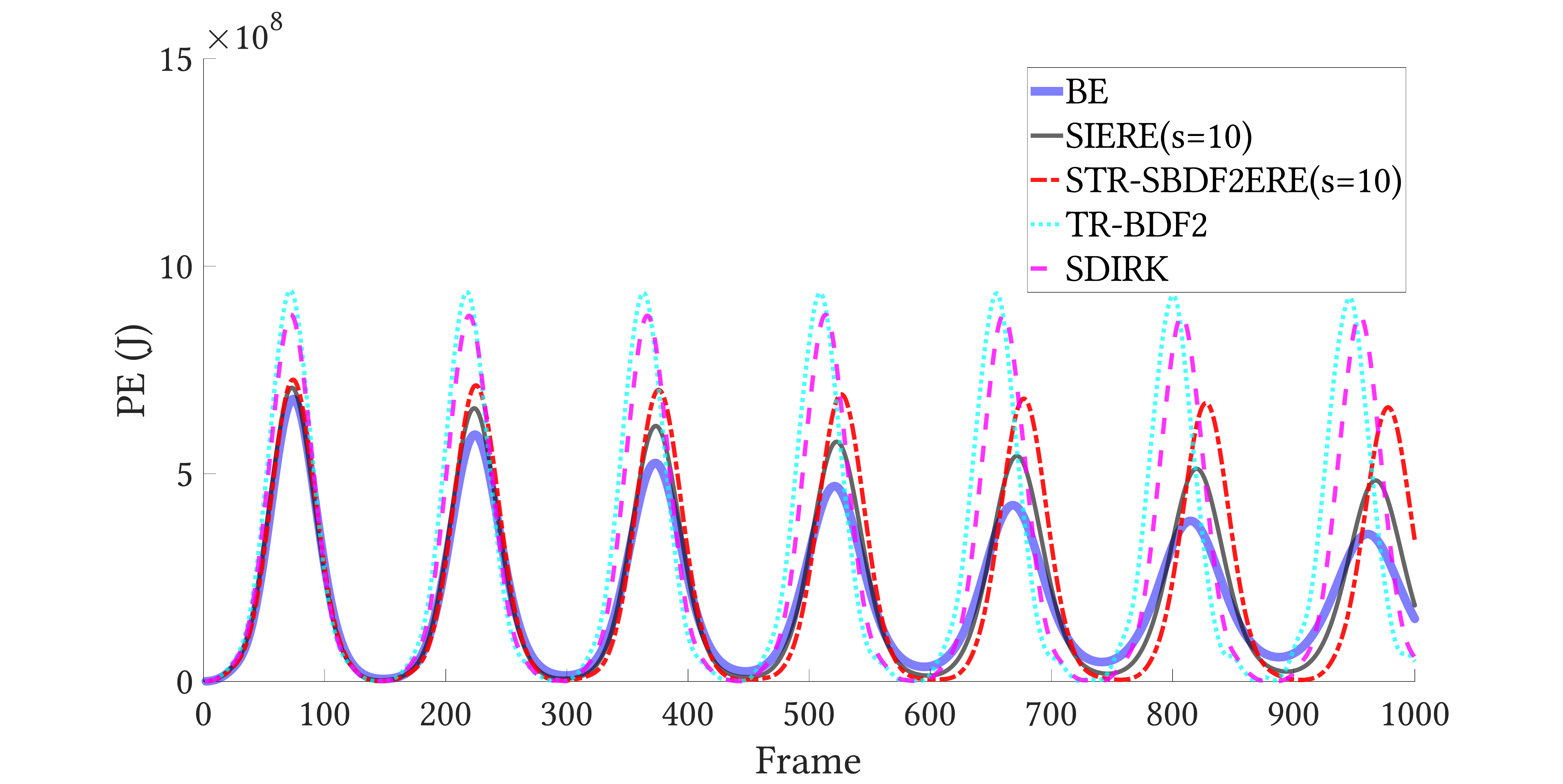}
		\put(25,30){\includegraphics[scale=0.08]{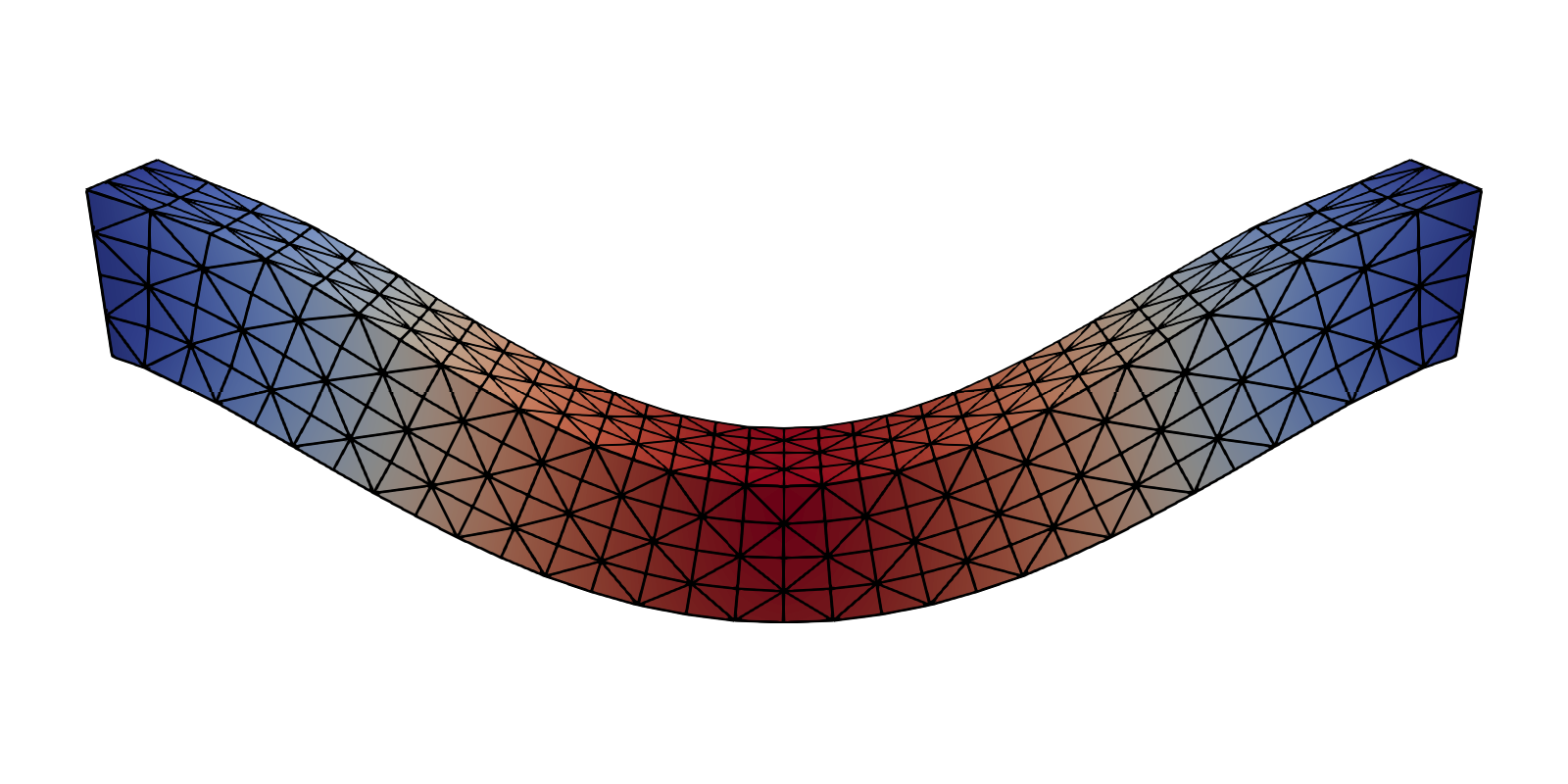}}
	\end{overpic}
	\includegraphics[trim=0 0 0 275, clip, width=5.0in]{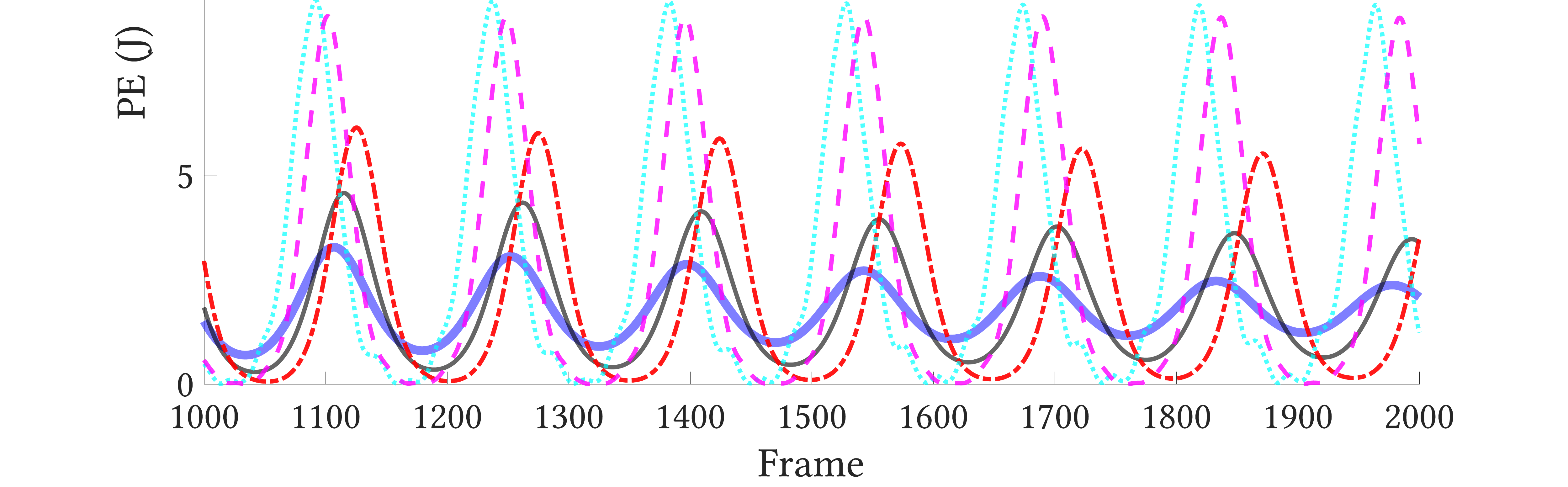}
	\caption{Potential energy plots for different integrators applied to a soft object: %body simulation: 
	BE (thick solid line), SIERE with $s=10$ (thin solid line), STR-SBDF2ERE with $s=10$ (dash-dot), 
	TR-BDF2 (dotted), and SDIRK (dashed). 
		A soft % cantilever 
		beam is fixed at its ends %on the two sides 
		and is subjected to gravity. 
		Notice that the TR-BDF2 and SDIRK energies do not decay by much, %preserve energy effectively, 
		whereas BE dissipates energy quickly. 
		SIERE %with ($s=10$) 
		is less damping %compared to 
		than BE but still much more damping than STR-SBDF2ERE, which in turn is still more damping than the two DIRK methods. 
		\label{fig:soft_energy}}
\end{figure}

Although SIERE is accurate for the first low frequency modes for a chosen $s$, 
its use of SI dampens  %is less accurate for 
the remaining eigenmodes significantly more than our %compared to 
DIRK methods. %, which have much less artificial damping than SI.
Hence, the DIRK methods have the potential to produce much livelier animations 
%are much more accurate for the entire spectrum of eigenmodes 
when applied to soft deformable objects. %problems.
Figure~\ref{fig:soft_energy} depicts  potential energy plots for a typical soft body animation problem, %likely to apper in computer graphics, 
where a homogeneous %cantilever 
beam that has been discretized with $5 \cdot (32\times4\times4) = 1920$ tetrahedral elements  is fixed at both ends
%at the two sides 
and is subject to gravity. 
The material parameters are chosen to be $E = 1\times1e5 \text{Pa}$ and $\nu = 0.4$, and the elastic force is given by %s are modelled by 
the nonlinear stable neo-Hookean model~\cite{smith17}.
We clearly see that the TR-BDF2 and SDIRK integrators correspond to much livelier animations compared to SIERE,
while SIERE in turn is still much less damping than BE.
 
%One might 
We may consider a method that is more accurate at the higher eigenmodes end than SI, but still uses ERE at the lower end of the spectrum.
One such integrator is STR-SBDF2ERE: it integrates the lower eigenmodes with ERE and the rest with STR-SBDF2, which is a semi-implicit version of TR-BDF2, given by
\begin{subequations}
\begin{eqnarray}
	\uu_{1/2} &=& \uu_0 + \frac 12 \left( I - \frac{h}{4} J_{F,0} \right)^{-1} \left( h \FF(\uu_0) \right)
	\label{eq:str-sbdf2-1}, \\
	\uu_1 &=& \uu_{1/2} + \frac 13 \left( I- \frac{h}{3} J_{H,1/2} \right)^{-1} \left( \uu_{1/2} - \uu_0 + h \bar{\FF} (\uu_{1/2}) \right), \quad\quad\\
	\text{where}\quad \bar{\FF}(\uu_{1/2}) &=& {\bf H}(\uu_{1/2}) + \phi_{1}(\frac h2 J_{G,1/2}) {\bf G}(\uu_{1/2}).
\end{eqnarray}
The Jacobian matrices are %and where 
$J_{F,0} = \frac{\partial \FF}{\partial \uu} (\uu_0)$, $J_{H,1/2} = \frac{\partial \bf H}{\partial \uu} (\uu_{1/2})$, and $J_{G,1/2} = \frac{\partial \bf G}{\partial \uu} (\uu_{1/2})$.
\label{26n}
\end{subequations}%
It can be seen from Figure~\ref{fig:soft_energy} that STR-SBDF2ERE conserves energy better %is more accurate 
than SIERE with the same value of $s$, but it still does not reach the liveliness of TR-BDF2 and SDIRK.
A fully implicit version of this integrator, TR-BDF2ERE might be a better alternative that is %both useful for stiff and soft problems.
useful for soft object animation problems.

%%%%%%%%%%%%%%%%%%%%%%%%%%%%%%%%%%%%%%%%%%%%%%%%%%%%%%%%%%%

\section{Contact and friction}
\label{sec:contact}   

%\uri{[Egor's example?]}   Follow \cite{li20}
%\input{friction_revise.tex}

In this section we consider situations that arise when the simulated soft object comes in contact with a surface. 
For instance, a soft robot's arm %toy bunny 
hits a wall or slides along some surface. 
The corresponding forces resulting from such events (corresponding to the ODE system being supplemented by equality and inequality algebraic constraints) 
can be written as the sum of contact and friction forces, $\ff_{\rm con} = \ff_{\rm c} + \ff_{\rm f}$ in~\eqref{1}.

%want to address the treatment of contact and friction in animation.
%We aim to simulate elastic bodies subject to frictional contacts, damping and external forces.

The question here is how $\ff_{\rm con}$ is related to nodal positions  \rev{$\qq$ and velocities $\vv$}.
Following the physical laws of contact and friction we can establish what these forces \emph{ought} to be.
Then, we derive approximations based on barrier and penalty methods in order to stay away from non-smooth or complex constraints  
in a manner that still yields high quality animations.
There is a significant amount of literature on this topic; see, e.g.,~\cite{kaufman08,daviet11,erleben17,verschoor19,geilinger20,larionov21,li20,Ferguson:2021}.
%Recently Li et al.~\cite{li20} have developed a method for efficiently resolving the original
%non-penetration constraint. 
Below we %will 
focus mostly on~\cite{li20}, since it produces versatile and
accurate results in the computer graphics context.
\rev{However, our description differs in some minor details}.
\rev{The methods discussed here are not similar to those in \cite{lotstedt,lp}.}

\subsection{Contact forces}
\label{sec:contactf}

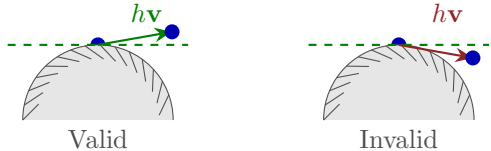
\begin{figure}
    \centering
    \begin{tikzpicture}[
interface/.style={
        % The border decoration is a path replacing decorator. 
        % For the interface style we want to draw the original path.
        % The postaction option is therefore used to ensure that the
        % border decoration is drawn *after* the original path.
        postaction={draw,decorate,decoration={border,angle=-45,
                    amplitude=0.3cm,segment length=2mm}}},
]

\tikzstyle{arrow} = [-{Stealth[scale=1,angle'=45]}, thick]

\def\grn{green!50!black}
\def\blu{blue!70!black}
\def\red{red!70!black}
\def\lightblu{blue!20!white}
\def\light{black!10!white}
\def\gray{black!70!white}
\def\offset{2}
\def\width{1}
\def\sqrttwo{1.414235624}

\begin{scope}[shift={(-\offset,0)}]

\fill[\blu] (0,0) circle (0.1);
\fill[\light] (225:\sqrttwo) arc (180:0:1);
\draw[\gray, interface] (225:\sqrttwo) arc (180:0:1);

% velocity
\draw[\grn, arrow] (0,0) -- (10:1);
\fill[\blu] (10:1) circle (0.1);
\node[\grn, anchor=south east] at (10:1) {$h \vv$};
\draw[\grn, dashed, thick] (-1.2,0) -- (1.2,0);
\node[\gray, anchor=north] at (0,-1) {Valid};
\end{scope}

\begin{scope}[shift={(\offset,0)}]
\fill[\blu] (0,0) circle (0.1);
\fill[\light] (225:\sqrttwo) arc (180:0:1);
\draw[\gray, interface] (225:\sqrttwo) arc (180:0:1);

% velocity
\draw[\red, arrow] (0,0) -- (-10:1);
\fill[\blu] (-10:1) circle (0.1);
\node[\red, anchor=south east] at (10:1) {$h \vv$};
\draw[\grn, dashed, thick] (-1.2,0) -- (1.2,0);
\node[\gray, anchor=north] at (0,-1) {Invalid};
\end{scope}

\end{tikzpicture}
    \caption{With large enough time steps, velocity based contact constraints
    may reject plausible steps. If a vertex in blue is constrained to have a
    strictly positive velocity with respect to the convex gray contact
    surface, then plausibly valid next-step configurations (right) may be
    erroneously rejected.} \label{fig:contact}
\end{figure}

Contact can be established by enforcing either a non-penetration constraint
\begin{subequations}
\begin{eqnarray}
\gb^{\text{np}}_\c(\qq) \geq \zero, \label{cona}
\end{eqnarray}
which prevents two objects from occupying the same space,
or alternatively using a velocity based constraint
\begin{eqnarray}
\gb^{\text{vel}}_\c(\vv) \geq \zero, \label{conb}
\end{eqnarray}
which restricts velocities to one side of the tangent half-space to the contact surface.
\label{con}
\end{subequations}%
In practice we may %often 
choose to linearize non-penetration constraints around $\qq_0$, e.g.,
\[
\gb^{\text{np}}_\c(\qq_1)
\approx  \gb^{\text{np}}_\c(\qq_0) + h\frac{\partial \gb^{\text{np}}_\c(\qq_0)}{\partial \qq} \vv_1 .
\]
%where superscript $\cdot^t$ fixes the configuration at time $t$ and $h$ is the size of the time step. 
However, this linearization may cause artifacts,
typically in highly curved areas and with large time steps; see
Figure~\ref{fig:contact}.
% shows an example where non-penetration based contact
%may produce a more plausible outcome.
 It is worth noting that the inequalities in the constraints \eqref{con} 
 %in this constraint
can be strict. % as in~\cite{li20}.

\begin{figure}
    \centering
    \includegraphics[width=0.5\textwidth]{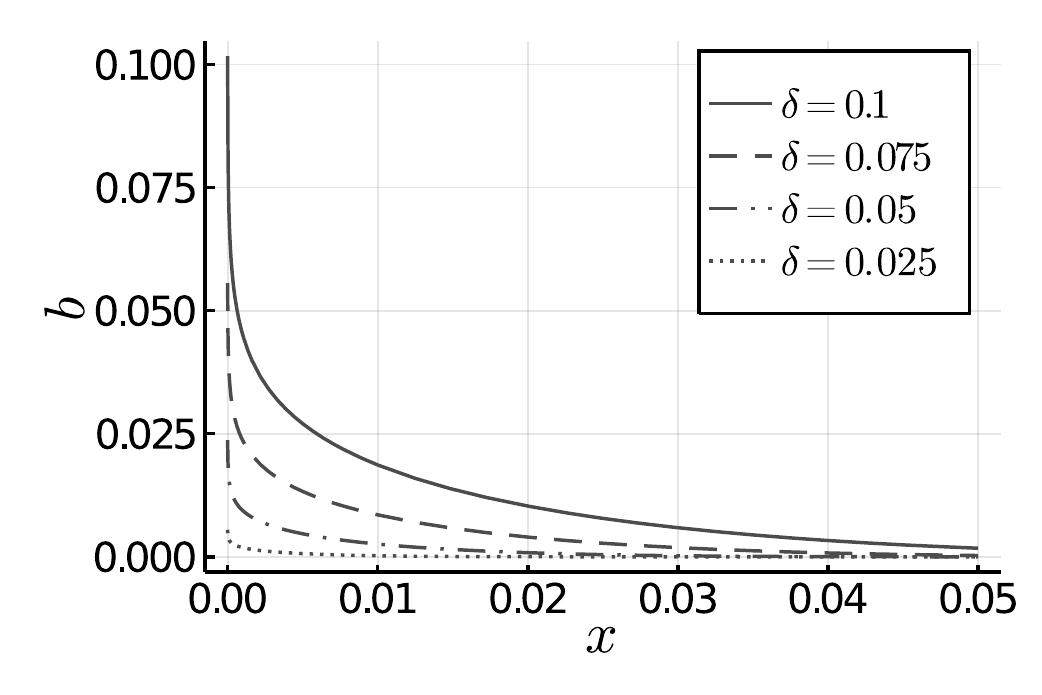}
    \caption{Barrier function \uri{$b = b(x; \delta)$} for different values of $\delta$. It is used in~\eqref{cforce} to approximate the contact force.}
    \label{fig:barrier}
\end{figure}

Let $\ContactDom$ denote the index set of contact nodes, and let $n_c \equiv |\ContactDom|$  %points. We  
count the total number of point contacts in the system. Then,
assuming we have a per contact distance (or ``gap'') function $\dd(\qq) \in \R^{n_c}$, we can construct the contact constraint. If the
distance function is signed, we can define $\gb_c^{\text{np}} \equiv \dd$ directly
\cite{larionov21} and rely on well-known techniques such as interior-point or
active-set methods~\cite{nw} for resolving inequality constraints. However, recent
works in the computer graphics context~\cite{geilinger20, li20} found value in formulating the
contact constraint as a penalty or barrier constraint explicitly. This approach has
%various 
several advantages:
\begin{itemize}
    \item The distance function can be unsigned to handle contact with objects of positive co-dimension~\cite{li21}.
    \item The contact constraint can be expressed as an equality constraint, thus simplifying the problem considerably. %simplifying the problem.
    \item Importantly and more specific to the current context, the entire simulation can be differentiated, which allows for more flexibility in higher level applications such as material parameter optimization~\cite{geilinger20}.
\end{itemize}
\rev{It's worth noting that this contact formulation can even be effective in rigid body dynamics \cite{Ferguson:2021}.}
It is worth noting that the function $\dd$ need not be a strict distance, but merely
continuous and monotonic. It
is even desirable for $\dd$ to be smooth~\cite{li20, larionov21},
whereas the true distance function can be non-smooth especially when dealing
with piecewise linear contact surfaces. To formalize the penalty based
contact constraint we could use a variety of functions such as soft-max or
truncated log-barriers. %We will 
Here we stick with a log-barrier $b$ with compact
support as in~\cite{li20} to enforce strict positivity on $\dd$, defining
%. We define the barrier by
\begin{align}
    b(x) = b(x; \delta)  \equiv \left\{ \begin{array}{ll}
    -(x-\delta)^2\log\left(\frac{x}{\delta}\right) & \text{if } x \in [0, \delta] \\
    0 & \text{otherwise}.
    \end{array}\right.
\end{align}
Figure~\ref{fig:barrier} shows how the barrier changes when $\delta$ is decreased. 
\rev{The barriers can be combined} over distances at each contact point in a straightforward manner 
\rev{\begin{align}
    b(\dd) \equiv \sum_{i=1}^{n_c} b(d_i),
\end{align}}%
where we have dropped $\delta$ for notational simplicity. 
We can now define the penalty force due to the barrier energy as
\begin{subequations}
\begin{eqnarray}
    \ff_\c(\qq) \equiv \blambda(\qq)^\top \frac{\partial \dd}{\partial \qq}(\qq), \label{cforcea}
\end{eqnarray}
where $\blambda(\qq) \in \R^{n_c}$ is the stacked vector of relative contact forces per contact point given by
\begin{eqnarray}
    \blambda(\qq)^\top = - \kappa\frac{\partial}{\partial \dd}b(\dd(\qq))  \label{cforceb}
\end{eqnarray}
for some conditioning parameter $\kappa > 0$. See \cite{li20} for
%\egorrev{%\sout{details on} 
\rev{the complete formulation of this barrier method
including setting $\delta$} and adaptively picking $\kappa$ for
efficient simulation.
This contact force then \egorrev{penalizes violations of} %\sout{enforces}} 
the equality constraint 
    \rev{\[b(\dd(\qq)) = 0\]}%
in place of \eqref{con}. \egorrev{Although this barrier formulation is motivated
by the interior-point method, it is distinct because here we are not
interested in converging to an arbitrarily small constraint violation.
Instead, the violation is bound to a physical quantity --- the distance
$\delta$ within which, objects appear to be in contact.}
\label{cforce}
\end{subequations}

\subsection{Friction forces}
\label{sec:friction}

To start we need to define a few quantities necessary to formulate the friction problem.
First we define the contact Jacobian $\Jac(\qq)$ as in~\cite{larionov21} to be a mapping from %our 
the velocity degrees of freedom $\vv$ to relative velocities at each contact point.
This quantity depends on the discretization of the contact surface. %In \cite{larionov21} 
We generate the surface using implicit functions, whereas~\cite{li20} 
use point-triangle and edge-edge proximity pairs to build the contact
Jacobian. Using the same notation as in \cite{larionov21} we also define the
change of coordinates matrices $\BB(\qq) = [\BB_\nml(\qq) |
\BB_\tng(\qq)]$ for all contact points, which give the normal and
tangential components of contact velocities respectively.
To further simplify notation we define the composition
\[
    \TT(\qq) \equiv \Jac(\qq)^\top\BB_\tng(\qq)
\]
to be the ``sliding basis''.
% Uri -- this ref is mentioned too many times....      as in \cite{li20}.

Now we can derive a friction force from first principles, namely the maximum dissipation principle (MDP). 
For each contact point $i \in \mathcal{C}$, MDP dictates that the friction force is defined by %as follows
\begin{eqnarray*}
\ff_i = \argmax_{\|\yy\| \leq \mu\lambda_{i}} -\vc_i^\top \yy , \label{eq:mdp}
\end{eqnarray*}
where $\vc_i \in \R^2$ is a relative tangential velocity at contact point $i$ and $\lambda_i$ is the
$i$th element of $\blambda$ as defined in \eqref{cforceb}.
This can be rewritten explicitly as %by
\rev{\begin{align}
\ff_{i} \in -\mu \lambda_{i} \left\{ \begin{array}{ll} \{\,\vc_i / \|\vc_i\|\,\} & \text{if } \|\vc_i\| > 0 \\
\{\,\vc_i \in \R^2 : \|\vc_i\| \leq 1\,\} & \text{otherwise} 
\end{array}\right.  .\label{eq:mdp_solved}
\end{align}}%
%We use $\|\cdot\|$ to mean the Euclidean norm. 
This relationship is commonly referred to as Coulomb friction. Ensuring this
inclusion strictly at the end of each time step has been a thorny point
in computer graphics simulation for some time
\cite{kaufman08,daviet11,erleben17,verschoor19,larionov21}. The non-smooth
transition at $\|\vc_i\| = 0$ calls for non-smooth optimization
techniques, which are not as well developed and are not as effective
when compared to their smoother counterparts.

\begin{figure}
    \centering
    \includegraphics[width=0.5\textwidth]{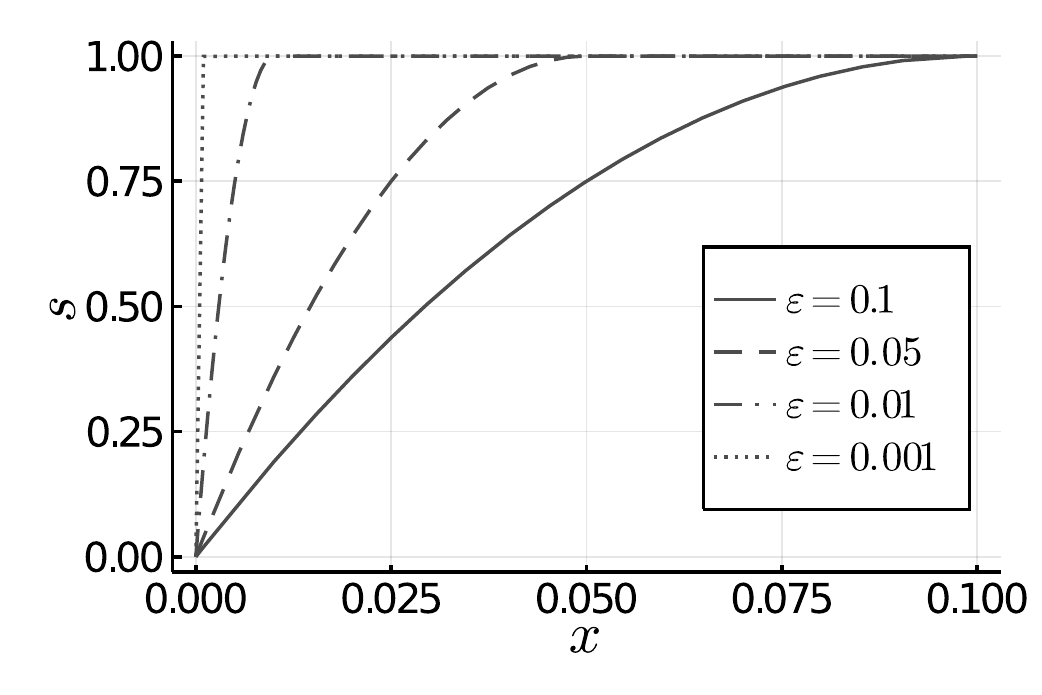}
    \caption{Plot of the smoothing function $s = s(x;\epsilon)$ for different values of $\epsilon$.
    It is used in~\eqref{fforce} through~\eqref{mdp_smoothed} to approximate the Coulomb friction force.}
    \label{fig:s}
\end{figure}

%In their recent graphics publication 
In the computer graphics context~\cite{li20} proposed to smooth this
transition at $\|\vc_i\| = 0$, which yields sufficient sticking (though
not absolute), given the short time periods used in typical %graphics
animations. Interestingly, older engineering works have also been
recommending this kind of
approximation~\cite{kikuuwe05,wojewoda08,awrejcewicz08} to improve hysteretic
behaviour and conveniently sidestep the numerical difficulties of Coulomb
friction.
The simplest smoothing redefines friction to have the form
\begin{subequations}
\begin{eqnarray}
    \ff_{i} = -\mu \lambda_{i} \nli(\vc_i) , \label{eq:mdp_smoothed}
\end{eqnarray}
where $\nli : \R^{2} \to \R^{2}$ defines the per-contact non-linearity
\begin{eqnarray}
    \nli(\vc_i) \equiv s(\|\vc_i\|)\left\{ \begin{array}{ll} \vc_i / \|\vc_i\| & \text{if } \|\vc_i\| > 0 \\
0 & \text{otherwise} 
\end{array}\right.
\end{eqnarray}
and the $C^1$ function $s(\cdot )$ defines the %smooth 
pre-sliding transition
\begin{eqnarray}
    s(x) = s(x; \epsilon ) \equiv \left\{\begin{array}{ll}
    \frac{2x}{\epsilon}-\frac{x^2}{\epsilon^2} & \text{if } x < \epsilon \\
    1 & \text{otherwise}.
    \end{array}\right.
\end{eqnarray}
%which 
This approaches the Coulomb friction as $\epsilon \to 0$. Figure~\ref{fig:s}
shows how %the 
\rev{this} function %$s (x)$ 
behaves when $\epsilon$ is decreased.
\label{mdp_smoothed}
\end{subequations}

To express the friction force as a function of all degrees of freedom, we
first need to define the non-linearity above for all stacked tangential
contact velocities $\vc \in \R^{2n_c}$:
\begin{align}
    \nl(\vc) \equiv (\nli(\vc_1), \nli(\vc_2), \dots, \nli(\vc_{n})) \in \R^{2n_c}.
\end{align}
For convenience let us define the diagonal matrix 
\[   \Lambda = {\rm diag}  [\lambda_1, \lambda_1, \lambda_2, \lambda_2, \ldots , \lambda_n,\lambda_n ] .\]
%consisting of 2x2 diagonal blocks $\lambda_iI_2$ as 
%\begin{align*}
 %   \Lambda \equiv \begin{pmatrix}
 %   \lambda_1I_2 & & \\ & \ddots & \\ & & \lambda_{n}I_2
 %   \end{pmatrix}
%\end{align*}
Then the total friction force can be written compactly as
\begin{eqnarray}
    \ff_\f(\qq,\vv) \equiv -\mu\TT(\qq)\Lambda(\qq)\nl\left(\TT(\qq)^\top\vv\right). \label{fforce}
\end{eqnarray}

Like with contact, the friction formulation here can deviate from traditional models, as allowed by the
limited simulation durations %of 
that arise in computer animation. \rev{This formulation deviates from \cite{li20} since
here we are not compelled to minimize an incremental potential. Instead we opt to solve a single system of non-linear equations.
We hope that this simple view of an otherwise complex problem is more accessible and can motivate novel efficient numerical methods with supporting analyses in future work.}

\rev{Above we have described a situation where discontinuities arise from
an interface of codimension 1. Cases with higher codimension can be handled in the same way upon viewing these objects as having some thickness, 
which is consistent with the assumption that contacting objects are allowed to rest at some distance apart. We refer the reader to 
\cite{li21}  for a more thorough analysis of additional difficulties encountered when simulating codimensional objects.}

\subsection{Combining the forces}
\label{sec:combine}

Putting together all forces for the equations of motion in \eqref{1} we get
\begin{align*}
    \dot{\qq} &= \vv , \\
    M\dot{\vv} &= -\frac{\partial W}{\partial \qq}(\qq) && \text{(elasticity)}\\
    &- D(\qq)\vv &&\text{(damping)}\\
    &+\blambda(\qq)^\top \frac{\partial \dd}{\partial \qq}(\qq) &&\text{(contact)}\\
    &-\mu\TT(\qq)\Lambda(\qq)\nl\left(\TT(\qq)^\top\vv\right) &&\text{(friction)} \\
    &+ \ff_{\ext}. && \text{(external forces)}
\end{align*}
The complexity of these equations gives rise to a plethora of potentially useful integration schemes
as discussed in previous sections. 
%For instance implicit Euler will produce
%\begin{align*}
 %   \genv{q}_+ &= \genv{q} + h\genv{v}_+ \\
  %  \genv{v}_+ &= \genv{v} + h\genv{M}^{-1}\big( -\frac{\partial W}{\partial \genv q}(\genv q + h\genv v_+) \\
  %  &- \genv{D}(\genv{q} + h \genv v_+, \genv{v}_+)\genv v_+ \\
  %  &+\contv{\lambda}(\genv q + h\genv v_+)^\top \frac{\partial \contv{d}}{\partial \genv q}(\genv q + h\genv v_+) \\
 %   &-\mu\physv{T}(\genv q + h\genv v_+)\Lambda(\genv q + h\genv v_+)f\left(\physv{T}(\genv q + h\genv v_+)^\top\genv{v}_+\right)  \\
  %  &+ \genv{f}_{\ext}\big).
%\end{align*}
However, even the simplest integrators face %this simple integrator poses 
challenges, since the friction term is often difficult to differentiate. 
It has long been known~\cite{desaxce98} that there is no single energy potential to derive these equations.
Nevertheless, a number of works in computer graphics have been able to incorporate friction
into a variational context~\cite{kaufman08,li20,larionov21}.

%%%%%%%%%%%%%%%%%%%%%%%%%%%%%%%%%%%%%%%%%%%%%%%%%%%%%%%%%%%

\section{Conclusion}
\label{sec:conclusion}

The task of animating flexible objects efficiently and reliably has many practical applications, yet it can be rather challenging.
Many  physics-based approaches have been considered in the past few decades, and more are sure to appear.
To more fully appreciate such efforts it is important to view resulting animation videos, not only formulas.
Fortunately, just about any paper in this area in recent years comes supplemented with such a video demonstration,
and the reader is thus urged to check them out. 
This area is teeming with \pai{opportunities} for applied and numerical mathematics research.

In this paper we have concentrated on the constrained, potentially very large dynamical system of differential equations
that results from an FEM discretization in space.
Aspects of numerical methods for \uri{large systems of} ODEs immediately arise, and the long Section~\ref{sec:siere} is devoted to this.
 Our new contribution here is the investigation of two popular DIRK methods,
\eqref{10} and \eqref{11}, which have been found to perform generally similarly with the exception that SDIRK can be {\em potentially} more efficient.
See in particular the new damping curves in Figure~\ref{fig:damp1}. 
Further quests into improving performance and utility using additive ODE methods have resulted in a model reduction scheme.
\rev{This gives rise, in particular, to the exciting possibility of using semi-implicit methods effectively, which eliminates the need for solving difficult
nonlinear algebraic equations.} 
In Section~\ref{sec:contact}, upon considering methods for handling contact and friction, we encounter large systems of ODEs with algebraic inequality
constraints, and these are handled by barrier methods that hail from the numerical optimization field.

%At 
In the same breath, it is imperative for numerical  and applied mathematicians
to keep in mind, as already mentioned in this paper's abstract, that the ``rules of the game'' are different here than in usual numerical analysis. 
\begin{itemize}
\item
The mathematical model is not fully given,
and it may well be adjusted in order to produce desired results. 
\rev{This implies, in particular, that the usual numerical analysis notion of comparing accuracy among methods is less relevant:
it is the ``eye norm'' that rules here.}
\item
Geometric numerical integration methods
are relatively scarce in this field even though the ODE problems in~\eqref{1ex} and~\eqref{1} look on a casual glance to be ripe for such a treatment.   
\item
The recent decisive employment of barrier penalty functions in Section~\ref{sec:contact} stems from the realization that, in the present context,
smoothness of the model is very important.
(NB The usual treatment of friction, in both graphics and engineering, previously considered it a non-smooth problem.)
Thus, the temptation to get closer to the exact non-smooth constraint can be harmful \pai{in these contexts}.
In other words, it's not merely a
question of ill-conditioning or of computational cost, as in the optimization literature. 
\item
Special FEM techniques had to be tailored for the
animation applications under consideration. %And so on.
\end{itemize}

There are several other aspects of the class of applications considered here that we have barely touched upon in our attempt to stay focussed,
and which require additional  expertise from new and traditional numerical and applied mathematics.
These include inverse problems, homogenization, various learning techniques, variational methods and FEM, and much more.

%\uri{ The paper ends here }

% You may incorporate your references as follows in your main tex file.
% Using BibTex is not recommended but can be handled.

%\begin{thebibliography}{99}

%\bibitem{A11}
%    \newblock  FirstNameInitial.  MiddleNameInitial. LastName, % first name middle initial. and then last name.  Only the first character in the paper title is capitalized.
%    \newblock Title of the paper,
%    \newblock \emph{Name of the Journal}, \textbf{Volume} (Year), StaringPage--EndingPage.

% Example of multiple authors:
%\bibitem{BFL} (MR1124979) [10.2307/2152750]
 %   \newblock Y. Benoist, P. Foulon and F. Labourie, % Use `and' connect the last two authors
 %   \newblock Flots d'Anosov a distributions stable et instable
 %    differentiables,
 %  \newblock (French) [Anosov flows with stable and unstable differentiable
 %    distributions], \emph{J. Amer. Math. Soc.}, \textbf{5} (1992), 33--74.

% Example of a preprint article with 7 digits archive number:
% \bibitem{quas}
% \newblock M. Entov, L. Polterovich and F. Zapolsky,
% \newblock Quasi-morphisms and the Poisson bracket,
%\newblock preprint, \arXiv{math/0605406}.

% Example of a book in the reference:
%\bibitem{rB} (MR1301779) [10.1007/978-1-4612-0873-0]
%    \newblock J.  Smoller,
%     \newblock \emph{Shock Waves and Reaction-Diffusion Equations},
%     \newblock 2$^{nd}$ edition,  Springer-Verlag, New York, 1994.

\bibliographystyle{plain}
\bibliography{ascher_sds20}

%\end{thebibliography}

%\medskip
% The data information below will be filled by AIMS editorial staff
%Received xxxx 20xx; revised xxxx 20xx.
%\medskip

\end{document}